\documentclass{aa} 
\usepackage{graphicx}
\usepackage{txfonts}
\usepackage{amsmath}
\usepackage[switch,mathlines]{lineno}
\usepackage[colorlinks,citecolor=blue,linkcolor=blue,urlcolor=blue]{hyperref}
\usepackage{multirow}
\usepackage{ulem}
\usepackage{comment}

\usepackage{adjustbox}

\newcommand{\fermi}{\textit{Fermi}\xspace}
\newcommand{\fermilat}{\textit{Fermi}-LAT\xspace}
\newcommand{\nustar}{\textit{NuSTAR}\xspace}

\newcommand{\swiftxrt}{\textit{Swift}/XRT\xspace}
\newcommand{\gm}{\ensuremath{\gamma}}
\newcommand{\pks}{PKS\,1830$-$211\xspace}

\usepackage{color}
\definecolor{tbd}{rgb}{0,0.6,0.3} 

\begin{document} 
   \title{High-energy variability of the gravitationally lensed blazar \pks}

   \author{
        Sarah M. Wagner
        \inst{1}\fnmsep\inst{2}\fnmsep\inst{3}
        \and
        Jeffrey D. Scargle\inst{4}
        \and
        Greg Madejski\inst{2}
        \and
        Andrea Gokus\inst{5}
        \and
        Krzysztof Nalewajko\inst{6}
        \and
        Patrick G\"unther\inst{1}
        \and
        Karl Mannheim\inst{1}
        }

   \institute{
        Julius-Maximilians-Universit\"at W\"urzburg, Fakult\"at für Physik und Astronomie, Institut für Theoretische Physik und Astrophysik, Lehrstuhl für Astronomie, Emil-Fischer-Str. 31, D-97074 W\"urzburg, Germany\\
        \email{sarah.wagner@uni-wuerzburg.de}
        \and
        Kavli Institute for Particle Astrophysics and Cosmology and SLAC National Accelerator Laboratory, Stanford University, Menlo Park, California 94025, USA
        \and 
        Institut de Física d’Altes Energies (IFAE), The Barcelona Institute of Science and Technology (BIST), E-08193 Bellaterra (Barcelona), Spain
        \and
        Astrobiology and Space Science Division, NASA Ames Research Center, Moffett Field, California 94035-1000, USA
        \and
        Department of Physics \& McDonnell Center for the Space Sciences, Washington University in St. Louis, One Brookings Drive, St. Louis, MO 63130, USA
        \and 
        Nicolaus Copernicus Astronomical Center, Polish Academy of Sciences, Bartycka 18, 00–716 Warszawa, Poland
        }
   \date{Submitted October 6, 2025}

  \abstract
   {The production site and process responsible for the highly variable high-energy emission observed from blazar jets are still debated. Gravitational lenses can be used as microscopes to investigate the nature of such sources.}
   {We study the broad-band spectral properties and the high-energy variability of the gravitationally-lensed blazar \pks, for which radio observations have revealed two images, to put constraints on the jet physics and the existence of a gravitationally-induced time delay and magnification ratio between the images.}
   {We utilize \swiftxrt, \nustar, and \fermilat observations from 2016 and 2019 to compare periods of low activity and high activity in \pks. 
   Short-timescale variability 
   is elucidated 
   with an unbinned power spectrum 
   analysis of 
   time-tagged \nustar photon data.
   To study the
   gravitationally-induced time delay in the \gm-ray light curve observed with \fermilat, 
   we  improve existing 
   autocorrelation function based 
   methods.}
   {Our modified auto-correlation method yields a delay of $t_0=21.1 \pm 0.1\,$d and magnification factor $a=0.13 \pm 0.01$. 
    These parameters remain time-invariant.
    In data from 2016 and 2019, the X-ray spectra remain remarkably stable, contrasting with extreme changes in \gm-rays.
    Both states can be fitted with a single component from Comptonisation of infrared emission from the dusty torus, with different \gm-ray states arising solely from a shift in the break of the electron energy distribution.
   }
  {The detection of a consistent lag throughout the whole light curve suggests that the \gm-rays originate from a persistent location in the jet.}

   \keywords{Galaxies: active, Galaxies: jets, Quasars: individual: \pks, Gravitational lensing: strong, Gamma rays: galaxies, X-rays: galaxies, X-rays: individuals: \pks, Acceleration of particles, Methods: data analysis}

   \maketitle


\section{Introduction}

Blazars are active galactic nuclei (AGN) with their jet aligned closely to our line of sight \citep{Blandford_2019, Urry_1995}.
The most energetic, steady radiation observed in the universe is identified with such sources \citep{MadejskiSikora_2016}. Extreme high-energy outbursts on time scales as short as days and shorter have been observed from blazars during the last decade, see e.g. \cite{Hayashida_2015}. However, the mechanisms producing this radiation and its characteristic variability, are still not fully understood.
The spectral energy distributions (SED) of blazars show a characteristic double humped structure. Due to the fact that the radiation associated with the low-energy peak shows polarization, it is expected to be produced via synchrotron emission from a relativistic electron distribution. 
The origin of the second hump in the SED is commonly modeled with inverse Compton (IC) scattering, i.e. the electrons up-scatter ambient photons to high energies. 
Available photon fields for this are either the previously produced synchrotron photons (synchrotron self Compton model, SSC) or external fields (external Compton model, EC) such as the broad line region, the accretion disk, the dust torus, or even other components of the jet. 
In addition to the electrons, \cite{Mannheim_1993} pointed out that coexisting protons could cause a contribution to the high-energy hump via synchrotron radiation or interactions with ambient photons.
Where exactly in the jet these processes take place is under debate since the large required energies hint at close vicinity to the central engine while arguments of photon self absorption suggest that the emission region is located further down the jet \citep{MadejskiSikora_2016}. 
\cite{Sikora_2009} showed that emission at distances of many parsecs from the jet core is possible and \cite{Nalewajko_2025} proposed that the necessary energy density compression could be achieved by tension of toroidal magnetic field during reconnection plasmoid mergers.

\subsection{Gravitational lensing}
The path of light can be affected by the curvature of space-time caused by very massive objects. 
If, for instance, a massive galaxy is located close to our line of sight to the source of interest, light that would have otherwise not reached us is bent towards Earth. 
Thus, the galaxy effectively acts as a lens which can increase the brightness and produce multiple images for the source of interest, in some instances even an Einstein ring. 
A more detailed explanation of the physics behind this process can be found in
\cite{Saha_2003} and \cite{Courbin_2002}.

Gravitationally-lensed sources are particularly interesting since properties of the lensing process induce a time delay that can constrain cosmological parameters such as the Hubble constant \citep[e.g., see the illustrative example of M\,87 by][and references therein]{Barnacka_2015}. Furthermore, emission and absorption features reveal information about the properties and structure of the lensing galaxy and the source of interest itself. 

\cite{Millon_2020} list time delay measurements for approximately 40 lensed systems based on optical monitoring. The delay is typically on the order of tens of days. More than 4500 strong gravitational lenses are known \citep{lenscat} and for about 220 of those the source is confirmed to be a quasar.\footnote{\href{https://research.ast.cam.ac.uk/lensedquasars/index.html}{https://research.ast.cam.ac.uk/lensedquasars/index.html}}
To date, only two \gm-ray bright blazars are known to be gravitationally lensed: \pks (see below) and B0218+357 \citep{Cheung_2014, Vovk_2016}. This work focuses on the former.

\subsection{PKS 1830-211}\label{sec:intro_1830}
As one of only two \gm-ray bright gravitationally lensed blazars confirmed to date, \pks is of particular interest in many areas of astronomy because multi-wavelength data can be used to study the lensing process, draw conclusions on cosmology and learn about the source as well as the lens itself.
This luminous flat spectrum radio quasar (FSRQ) is located at a redshift of $z=2.507$ \citep{Lidman_1999}.
A detailed literature review on the source is provided in \cite{Wagner_phd}.
Radio observations reveal two lensed images of the source which potentially include contribution from an extended jet \citep{PrameshRao_Subrahmanyan_1988, Subrahmanyan_1990, Jauncey_1991, vanOmmen_1995, Patnaik_Porcas_1996}.
Most recently \cite{Muller_2020} detected a third image with ALMA data but since this is about 140 times dimmer it is neglected for the purpose of this study.

While the main lensing galaxy is located at $z_{\text{lens}} = 0.886$ \citep{WiklindCombes_1996}, observations with the Hubble Space Telescope (HST) revealed that the field of \pks contains several other galaxies \citep{Lehar_2000}.
Hence, the exact lens structure remains elusive but the data can be well modeled with one or two mass distributions described by a singular isothermal sphere (SIS) \citep[CASTLES\footnote{~\href{https://lweb.cfa.harvard.edu/castles/}{https://lweb.cfa.harvard.edu/castles/}, CfA-Arizona Space Telescope LEns Survey of gravitational lenses}]{Falco_1999,Lehar_2000}.

The main images of the blazar \pks can also be resolved in X-rays at the resolution limit $(\approx 1'')$ of the \textit{Chandra} telescope \citep{deRosa_2005}. The images can not be resolved with current \gm-ray instruments, 
causing the lensed blazar to appear as one point source. 

The \textit{observed} light curve is a superposition of two copies of the \textit{intrinsic} light curve of the source,
\begin{linenomath*}
    \begin{equation}
        y(t) = x(t) + a \, x(t-t_0) \,.
        \label{eq:lc_model}
    \end{equation}
\end{linenomath*}
where the second term is the copy with relative lag $t_0$ and magnification ratio $a$ defined relative to the normalized reference copy $x(t)$.
The parameters $t_0$ and $a$ are referred to as \textit{lens observables} throughout this work.

\pks was not detected in very high energies \citep{HESS_2019} so far.
Most recently \cite {Vercellone_2024} presented a broad band SED analysis of the source.\\

\begin{figure*}[t]
    \centering
    \includegraphics[width=\linewidth]{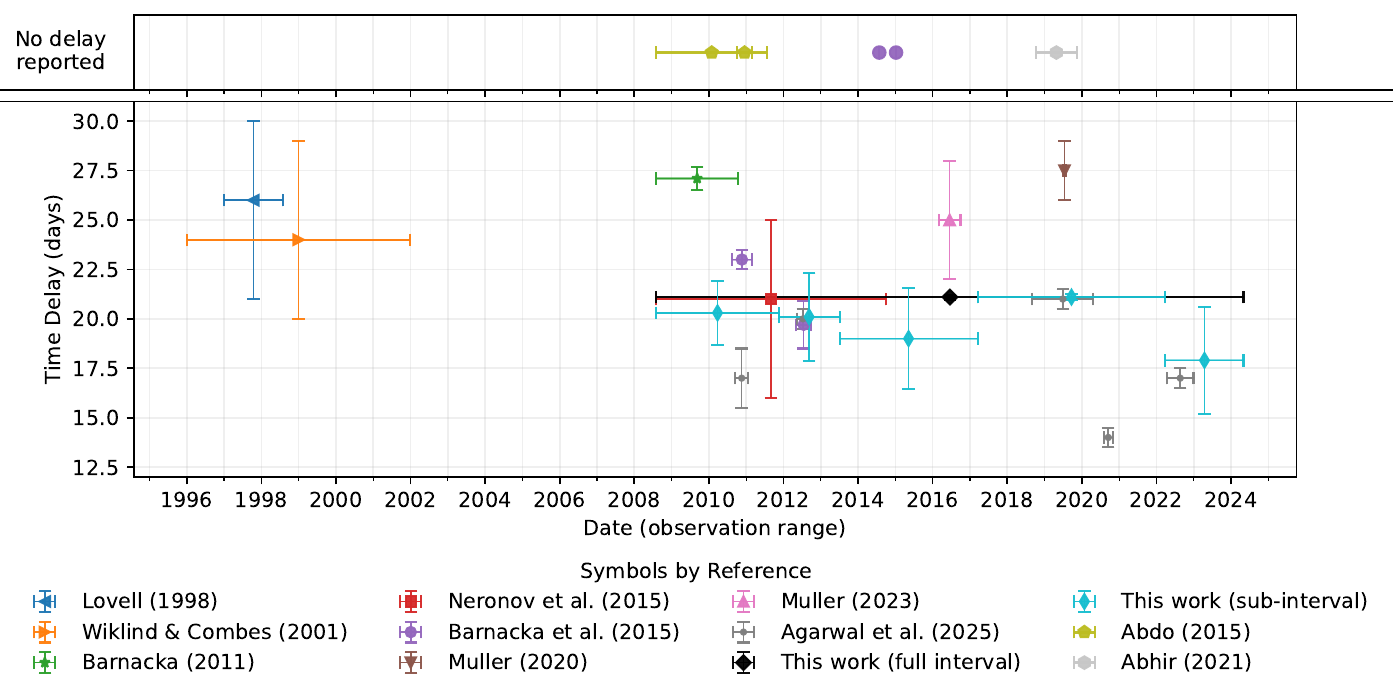}
    \caption{Literature values for the delay between the two major images of \pks.
    Triangle markers are based on radio data while the rest is based on \fermilat. 
    The result of \cite{vanOmmen_1995} at $44 \pm 9$ for 1990 Jun - 1991 Jul and the additional delay of $76^{+25}_{-15}$ found in \cite{Neronov_2015} are not shown. }
    \label{fig:lit_delay}
\end{figure*}

Although the time delay $t_0$ has been extensively studied, there is no clear consensus on its value nor on its existence. In Fig.\,\ref{fig:lit_delay} we present results published in the literature. 
We elaborate the methods relevant to this work in Sec.\,\ref{sec:lag} and go into more detail in Appendix\,\ref{sec:gaps} but refer the reader either to a more detailed summary in \cite{Wagner_phd} or to the original publications for full context. 

\cite{vanOmmen_1995} and \cite{Lovell_1998} utilize radio observations to model two radio cores and additional components.
The flux of the radio cores is used to determine the delay and accounting for the contribution of a potential Eisntein-ring explains differences between the obtained results. 
Similarly, the flux of molecular absorption lines in each image can be used to determine the delay \citep{WiklindCombes_2001}. 
Based on more recent ALMA data, \cite{Muller_2020} obtain a delay of 26-29\,d based on a lensing model and the adopted Hubble constant and \cite{Muller_2023} use monitoring to get $25\pm3$\,d.

In \gm~rays, the following time series analysis methods were used to determine the delay of the superimposed \fermilat light curve (Eq.\,\ref{eq:lc_model}).
The auto-correlation (ACF) or specifically the discrete auto-correlation (DCF) is often used albeit with inconclusive results. 
\cite{Barnacka_2011} introduced the ``double power spectrum'' method (DPS), where the power spectrum of the light curve is analyzed for a sinusoidal modulation that depends on the lens parameters, see Sec.\,\ref{sec:lag}.
\cite{Abdo_2015} made use of the wavelet transform and \cite{Neronov_2015} utilize the structure function, which is closely related to the ACF. 
The latter were the only ones to point out an additional larger delay explained by microlensing. 
\cite{Barnacka_2015} and \cite{Agarwal_2025} divide the light curve in individual segments, as done in this study as well, and find slightly varying delays through application of several methods. 
It becomes clear that the typically utilized methods are not equally suited to detect a delay in the noisy and unevenly binned light curve data of \fermilat.\\

Based on radio observations which resolve the two lensed images, \cite{PrameshRao_Subrahmanyan_1988} and \cite{Nair_1993} showed that the flux magnification ratio varies with time and frequency.
\cite{vanOmmen_1995}, \cite{Lovell_1998}, \cite{Jin_2003}, \cite{Marti-Vidal_2013}, \cite{Muller_2020}, and \cite{Muller_2023} find a magnification ratio slightly larger than one. 
\cite{Marti-Vidal_2020} directly fitted for the flux-density ratio in ALMA observations and show that it varies between 1.2 and 1.5 and shows frequency dependent behaviour (in radio) during the first half of 2012. 
They argue that the size of the core, which is responsible for the radio emission, is too large to explain the changing flux ratio of the lensed images with micro-lensing. Instead they propose that the frequency dependence of the flux ratio is related to opacity effects close to the base of the jet. We note, however, that this is not directly transferable to the high-energy regime which is discussed in this work.


\section{Data}
\label{sec:data}

\begin{figure*}[t]
    \centering
    \includegraphics[width=\linewidth]{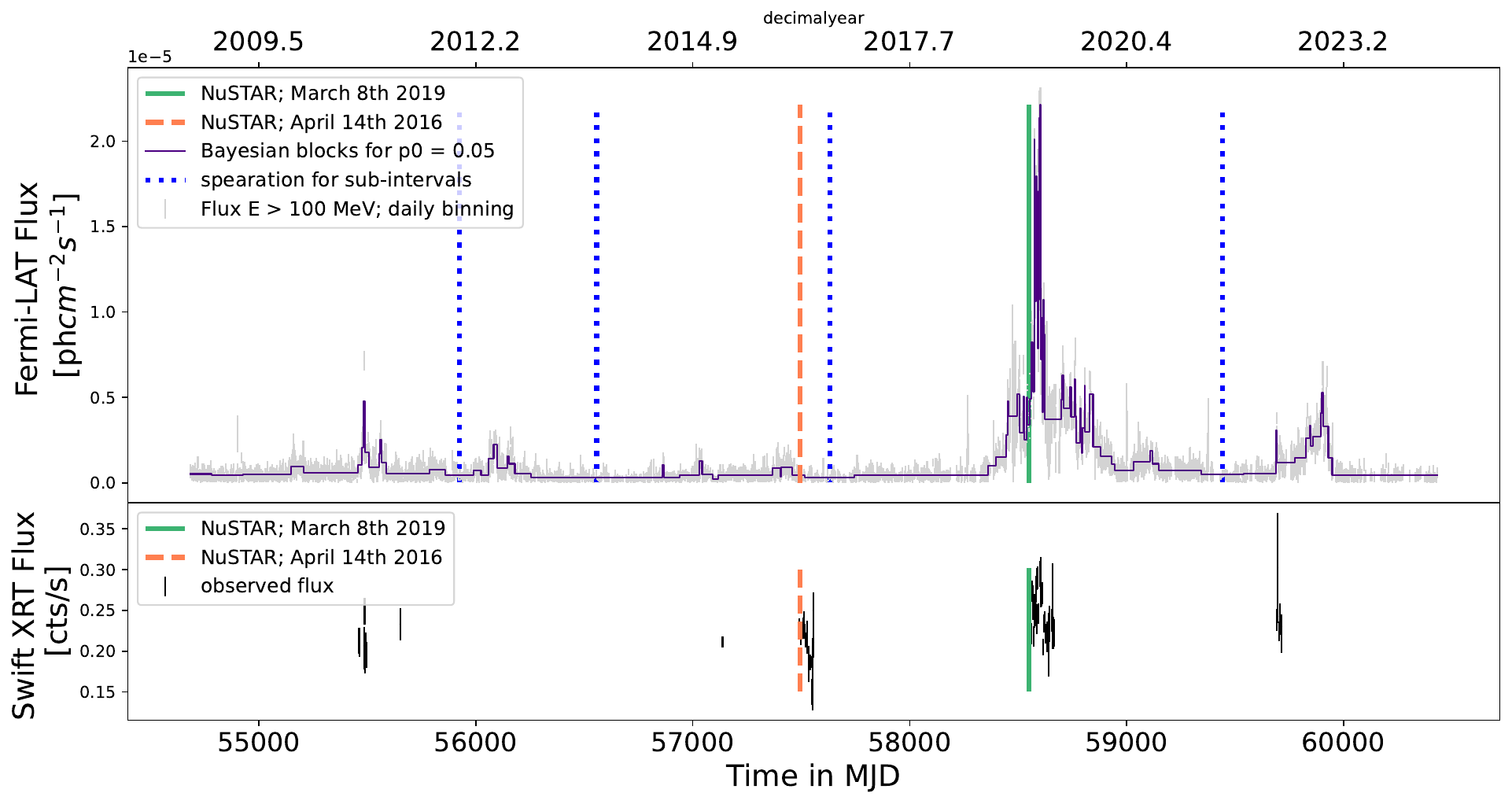}
    \caption{Light curve of \pks in daily binning with an orange and a green vertical line indicating the two available \nustar observations. For the \fermilat data (top) we neglect bins with TS $ < 0$, bins with uncertainties ten times larger than the flux, and bins affected by solar activity (see \ref{sec:data_fermi}). The best piece wise constant step function is shown in the middle panel for a false alarm probability of 5\% (p0 = 0.05) and blue vertical dotted lines indicate the division into sub-intervals (1-5). For the \swiftxrt data (bottom) we also omit upper limits}
    \label{fig:lcs}
\end{figure*}

\subsection{\swiftxrt}
The X-Ray Telescope onboard the Neil Gehrels Swift Observatory \citep[\swiftxrt, ][]{Gehrels_2004} performed a total of 90 observations of \pks to this day. 
This data can be accessed through the automated data processing website\footnote{\href{http://www.swift.psu.edu/monitoring/}{http://www.swift.psu.edu/monitoring/}}.
The light curve is shown in the bottom of Fig.\,\ref{fig:lcs}.
For the quasi-simultaneous spectral analysis, we consider the \swiftxrt observations listed in Tab.\,\ref{tab:swift_obs},\footnote{~Bold numbers are used as ID reference} which were performed before and after the two \nustar observations on 2016 April 14 and 2019 March 8. 
\begin{table}[h]
\centering
\caption{\swiftxrt observations for quasi-simultaneous spectral analysis (pc = photon-counting, wt = windowed-timing)}
\begin{tabular}{l|c|c}
Date & ObsID & Mode  \\
    \hline
    \hline
    2016 Apr 12 & 000812220\textbf{02} & pc + wt \\
    \hline
    2016 Apr 19 & 000384220\textbf{16} & pc \\ 
    \hline
    2019 Mar 06 & 000384220\textbf{36} & pc \\ 
    \hline
    2019 Mar 09 & 000384220\textbf{37} & pc \\ 
    \end{tabular}
    \label{tab:swift_obs}
\end{table}
For consistency, we only consider data taken in photon-counting mode (pc), although there is data available in windowed-timing mode (wt) for one of the observations. We will use the last two digits of the ID to refer to each corresponding \swiftxrt observation. 
We downloaded the data from \texttt{HEASARC}\footnote{\texttt{HEASARC} Data Archive, \href{https://heasarc.gsfc.nasa.gov/cgi-bin/W3Browse/w3browse.pl}{https://heasarc.gsfc.nasa.gov/cgi-bin/W3Browse/w3browse.pl}} and processed them with the standard procedure for photon-counting mode (see XRT User's Guide\footnote{XRT User's Guide, \href{https://swift.gsfc.nasa.gov/analysis/}{https://swift.gsfc.nasa.gov/analysis/}}) using the calibration database (CALDB 20211105) and \texttt{xrtpipeline} (version 0.13.7). We define a circular source region with radius of $45''$ centered on the source
as well as a circular background region with radius of $180''$ in a field that does not contain the source for each observation. Based on this, we extract source and background spectra in \texttt{XSELECT}\footnote{XSELECT Home Page, \href{https://heasarc.gsfc.nasa.gov/docs/software/lheasoft/ftools/xselect/index.html}{https://heasarc.gsfc.nasa.gov/docs/software/ lheasoft/ftools/xselect/index.html}} setting the event grade filter to 0-12 which is standard for photon counting mode.
Ancillary response files (arf) centered on the position of the source are created based on the exposure file for each observation with \texttt{xrtmkarf} and manually added to each spectrum with \texttt{grppha} along with the corresponding background spectrum and
the latest response matrix file (\texttt{swxpc0to12s6\_20130101v014.rmf}) from \texttt{CALDB}.
The spectra are grouped with \texttt{grppha} to ensure each bin contains at least one count. For spectral fits based on this data we consider the energy range between 0.5 and 10\,keV while ignoring bad channels.

\subsection{\nustar}
\label{sec:data_nustar}
The Nuclear Spectroscopic Telescope Array \citep[\nustar, ][]{Harrison_2013} is a focusing high-energy X-ray telescope operating from 3 to 79\,keV. In addition to an archival \nustar observation of \pks from 2016 April 14 (ObsID: \textbf{6016}0692002, exposure $\approx$ 22\,ks), we present our ToO observation from 2019 March 8 (ObsID: \textbf{8046}0628002, exposure $\approx$ 41\,ks). We will refer to each \nustar observation with the first four bold digits of its observation ID. 

We reduce and extract data of the two Focal Plane Modules A and B (FPMA and FPMB) with the standard methods\footnote{\nustar Data Analysis guide \href{https://heasarc.gsfc.nasa.gov/docs/nustar/analysis/}{https://heasarc.gsfc.nasa.gov/docs/ nustar/analysis/}} using \texttt{NUSTARDAS} (version v2.1.2) distributed in \texttt{HEASOFT} and the calibration database (CALDB) 20211202 via \texttt{nupipeline} (version~0.4.9).  This results in generation of cleaned event files, which contain information about the position (on the detector), the time stamp, and a Pulse Height Analysis (PHA) of all events detected during the \nustar pointing.  

For the spectral analysis, the spectra and response files are created with \texttt{nuproducts} for a circular source region of radius $50''$ and a circular background region of radius $120''$ again centered on and off the source, respectively. The obtained spectra are re-binned with \texttt{grppha} to ensure at least one count per bin. For both spectra, we consider the data in the energy range of 3 - 50 keV, where the source is clearly detected. 

Although the \nustar measurements are relatively brief, we search for significant variability within each observation (see Sec.\,\ref{sec:nustar_variab}.) based on unbinned photon data. These data are obtained by generating a new event file using  \texttt{XSELECT} where we select photon events with raw PHA channels 36 - 1709 inclusively (corresponding, roughly, to the energy range of 3 - 50 keV), apply an event grade filter 0-12, and filter events only from the source region - again a circle with radius $50''$;  we do not apply any additional binning. We subsequently convert this event file to an ascii file through \texttt{fdump}. It contains information about the arrival time, the event grade, and the position of each detected photon. This is done individually for FPMA and FPMB for both observations. As an added test of our analysis, we also extracted the data from the $120''$ region in the center of the Perseus Cluster, which although it includes flux from the central AGN 3C84, is dominated by the non-variable thermal cluster emission \citep[see, e.g.,][]{Rani_2018}.

\subsection{\fermilat}
\label{sec:data_fermi}
The Large Area Telescope on board the \fermi~satellite is a pair-conversion telescope operating from 20\,MeV up to 1\,TeV \citep[\fermilat, ][]{Atwood_2009}. Its setup allows for continuous observations over 15 years and is therefore uniquely suited for studying temporal behaviour such as the time delay in \pks. 

The light curve is computed with daily binning from 2008 August 5 to 2024 May 3 (MET 239587201 - 736387205) for an energy range between 100\,MeV and 300\,GeV. We performed the standard data reduction process\footnote{\fermilat Data Analysis Documentation, \href{https://fermi.gsfc.nasa.gov/ssc/data/analysis/documentation/}{https://fermi.gsfc.nasa.gov/ssc/data/analysis/documentation/}} with \texttt{Science Tools 1.2.23} and \texttt{fermipy 0.20.0} \citep{fermipy}, and extracted all events suitable for a scientific analysis\footnote{SOURCE class events, zenith angle $z\leq90$, and \texttt{(DATA\_QUAL>0)\&\&(LAT\_CONFIG==1)}}.

We model all sources included in the \fermilat Fourth Source Catalog \citep[4FGL;][]{4fgl} that are within a region of interest (ROI) of $15^\circ$ around the source, and model the isotropic diffusion emission with \texttt{iso\_P8R3\_SOURCE\_V2\_v1} and the Galactic diffuse emission with \texttt{gll\_iem\_v07}. The used response function is \texttt{P8R3\_SOURCE\_V2}, which was provided post-launch.
The data are modelled and optimised following a maximum likelihood approach. The significance of the \gm-ray emission for each model component is determined by comparing the likelihood function $\mathcal{L}$ for a model with and without the respective source, and given through the test statistic $\mathrm{TS}=2\Delta\mathrm{log}(\mathcal{L})$ \citep{mattox}.

We model each source with the best-fit spectral type according to the 4FGL, except for \pks, for which we employ a power law instead of a log parabola, since the spectral curvature is not significant at daily binning for most of the time.
For each light curve bin, we keep the spectral parameters for \pks free, as well as the normalisation for all sources within $3^\circ$ around that source plus sources that have a TS value of $>$ 500. 
The resulting light curve contains 5750 bins but for 127 of them the fit did not converge so we neglected these points \citep{LCR_2023}.

The above-mentioned TS value is utilized to quantify the detection significance of \pks in each bin of the light curve. There are different methods to deal with the inclusion or exclusion of bins associated with a low detection significance. Common procedures are to neglect them, or to replace them with an estimated flux such as the minimum flux in the light curve or a value below the upper limit \citep[see, e.g.,][]{Gokus_2021}. 
Relatively large  test statistic cutoffs are appropriate when searching for previously unknown sources, but for known sources including all data provides the most information.
Throughout this paper filtering with TS $>0$ reflects this choice and is referred to as TS filter (TSF). Negative TS values are not meaningful but can result from the fit parameters reaching the limits of their allowed intervals without having maximized the likelihood profile \citep{LCR_2023}.

To ensure data quality and reliable flux value estimates, we require for each bin of the light curve:
(1) TS $>$ 0, 
(2) flux $>$ flux error \citep{Meyer_2019},
(3) at least 3 predicted photons,
(4) no contamination by the sun, i.e. angular separation $>$ 5 degrees if there are more than $10^{8}$ counts per day reported in the \fermi-GBM solar flare catalog\footnote{~Although the solar flare catalog of the \textit{Fermi} Gamma-ray Burst Monitor (GBM) covers an energy range of 8 keV to 40 MeV, we take this to provide a sufficient estimate for the flux at higher energies in this context \href{https://hesperia.gsfc.nasa.gov/fermi_solar/}{https://hesperia.gsfc.nasa.gov/fermi\_solar/}}.
This leaves 3897 bins (67.8\%), which are shown in the top panel of Fig.\,\ref{fig:lcs}.
Additionally, we generate an imputed, evenly binned light curve via linear interpolation which is used for the Metric Optimization analysis.

The best piece wise constant representation of this light curve is computed with the Bayesian block algorithm \citep{Scargle_2013}\footnote{computed  for \textit{point measurements} under the assumption that the flux errors follow a Gaussian distribution, which is valid when there are a large number of photon counts in a given bin.
When photon counts are low, however, the errors become more Poissonian, resulting in smaller error values for those bins.} 
and shown in the top panel of Fig.\,\ref{fig:lcs}. 

We studied the impact of different priors on the Bayesian blocks and found that at a false alarm probability of $p_0=0.05$ the number of blocks is fairly stable and most variations seem to be captured without over-fitting the data.

Five sub-intervals were chosen to study whether the lens observables vary over time (see Sec.\,\ref{sec:lag}) based on time ranges of significant large-scale flaring.
These times were chosen to be the center of the lowest block separating the main flare peaks \citep[HOP algorithm, see]{Wagner_2021} as shown by the vertical dotted lines (MJD 55924.74, 56556.99, 57631.49, 59441.24) in the top panel of Figure \ref{fig:lcs}.
This partitions the full interval into non-overlapping sub-intervals without loss of data.

In addition to the light curve, we also compute two LAT spectra for the SED modeling of the source. We use the same infrastructure, selection parameters, and models for two time ranges that we have selected based on the availability of hard X-ray data from \nustar. 
Based on the Bayesian blocks, we identified intervals of relatively constant flux, and chose those blocks covering the \nustar observing dates as the time ranges for the analysis of the \fermi-LAT data. For the year 2016, we compute a spectrum from MJD 57459 (2016-03-12) until (incl.) MJD 57516 (2016-05-08); for the year 2019, the time range for the spectrum is from MJD 58544 (2019-03-02) until (incl.) 58553 (2019-03-11).\footnote{The 2019 spectrum covers a shorter interval due to the enhanced \gm-ray activity, whereas the 2016 spectrum spans over a longer, more constant period (from Bayesian blocks).}
While the modelling approach is the same, we free all spectral parameters of sources within $3^\circ$ of PKS\,1830$-$211 during the fit, and the normalisation parameter for sources within $5^\circ$ of PKS\,1830$-$211.


\section{X-ray Spectral Analysis}
\label{sec:xray_spec}

For the available X-ray data, we perform independent (see Appendix\,\ref{sec:xray_spec_independent}) as well as joint spectral fits  in \texttt{XSPEC}\footnote{\href{https://heasarc.gsfc.nasa.gov/xanadu/xspec/}{https://heasarc.gsfc.nasa.gov/xanadu/xspec/}} (Version 12.12.1) using an absorbed powerlaw ($N(E)\propto E^{-\Gamma}$).
The absorption is modeled with \texttt{tbabs} \citep{Wilms_tbabs} where the Galactic H$_\text{I}$ column density is in one run free to vary and in another fixed to the galactic value N$_{\mathrm{H}}~=~0.188~\cdot~10^{22}\,\mathrm{cm}^{-2}$  \citep{HI4PI_2016}.
The flux is determined with \texttt{cflux}.
While the fit statistic is computed with Cash statistics \citep{Cash_79}, we assess the goodness of each fit with $\chi^2_C$ computed with the Churazov weighting and normalized by the degrees of freedom (dof).
We also account for potential absorption in the lensing galaxy with \texttt{ztbabs} \citep[version 2.3,][]{Wilms_tbabs}.

In order to compare the overall X-ray flux of \pks in 2016 and 2019, we perform a simultaneous analysis of each \nustar observation with the previous and subsequent \swiftxrt observations.
Analogous to the individual fits (see Appendix\,\ref{sec:xray_spec_independent}) we use an absorbed power-law accounting for Galactic absorption, i.e. the model \texttt{constant$\,\times\,$tbabs$\,\times\,$ cflux(powerlaw)} where the normalization of the powerlaw and the constant for FPMA is set to one and the other constants account for differences in calibration ($\approx 1.04$ for FPMB and $\approx$ 0.45 for each \swiftxrt observation). We compute the flux in the range from 0.5 to 50\,keV. Our best fit parameters are listed in Tab.\,\ref{tab:joint_1_spec}. 

Additionally, we fit a power-law that is additionally absorbed by the lensing Galaxy ($z=0.88$, see Sect.\,\ref{sec:intro_1830}), 
i.e. \texttt{constant$\,\times\,$ztbabs$\,\times\,$tbabs$\,\times\,$ cflux(powerlaw)} and show the results in Tab.\,\ref{tab:joint_2_spec}

\begin{table}[]
    \centering
    \caption{Joint \nustar and \swiftxrt spectral analysis best fit parameters for \texttt{constant$\,\times\,$tbabs$\,\times\,$ cflux(powerlaw)} with the power-law index $\Gamma$ and the Galactic H$_\text{I}$ column density $n_H$ free to vary (top rows) and fixed to the Galactic value (bottom rows) as well as the estimated flux $F_{0.5-50}$ within 0.5-50\,keV in $10^{-12}$ erg/(cm$^2$s) for the 2016 data (IDs: 02 + 16 + 6016) and the 2019 data (IDs: 36 + 37 + 8046).}
    \resizebox{\linewidth}{!}{
    \begin{tabular}{c|c|c|c|c}
        yy & N$_{\mathrm{H}}$ $\left[\frac{10^{22}}{\text{cm}^{2}}\right]$ & $\Gamma$ & $\chi^2_C$/dof & $F_{0.5-50}$ $\left[10^{-12} \frac{erg}{cm^2s}\right]$ \\
        \hline
        \hline
        16 & $1.06 \pm 0.065$ & $1.468 \pm 0.016 $ & 0.92 & $ 53.81 \pm 0.78 $ \\ \hline 
        19 & $0.996 \pm 0.082$ & $1.466 \pm 0.012 $ & 0.92 & $ 50.54 \pm 0.55 $ \\ \hline 
        \hline
        16 & $0.188$ & $1.372 \pm 0.015 $ & 1.35 & $ 55.67 \pm 0.83 $ \\ \hline 
        19 & $0.188$ & $1.414 \pm 0.011 $ & 1.19 & $ 51.00 \pm 0.56 $ \\ 
    \end{tabular}
    }
    \label{tab:joint_1_spec}
\end{table}

\begin{table*}[h]
    \centering
    \caption{Joint \nustar and \swiftxrt spectral analysis best fit parameters for \texttt{constant$\,\times\,$ztbabs$\,\times\,$tbabs$\,\times\,$ cflux(powerlaw)} with the power-law index $\Gamma$ and the Galactic H$_\text{I}$ column density N$_{\mathrm{H}}$ fixed to the Galactic value, the column density zN$_{\mathrm{H}}$ in the lensing galaxy at $z=0.88$, and the estimated flux $F_{0.5-50}$ within 0.5-50\,keV in $10^{-12}$ erg/(cm$^2$s) for the 2016 and 2019 data (see left column). These parameters are used for further analysis and illustrated along with the data in Fig.\,\ref{fig:spec}.}
    \begin{tabular}{c|c|c|c|c|c|c}
        year & IDs & N$_{\mathrm{H}}$ $\left[\frac{10^{22}}{\text{cm}^{2}}\right]$ & zN$_{\mathrm{H}}$ $\left[\frac{10^{22}}{\text{cm}^{-2}}\right]$ & $\Gamma$ & $\chi^2_C$/dof & $F_{0.5-50}$ $\left[10^{-12} \frac{erg}{cm^2s}\right]$ \\
        \hline
        \hline
        2016 & 02 + 16 + 6016 & 0.188 & $4.29 \pm 0.33$ & $1.479 \pm 0.016 $ & 0.92 & $ 53.89 \pm 0.77 $ \\ \hline 
        2019 & 36 + 37 + 8046 & 0.188 & $3.91 \pm 0.41$ & $1.476 \pm 0.013 $ & 0.91 & $ 50.56 \pm 0.54 $ \\  
    \end{tabular}
    \label{tab:joint_2_spec}
\end{table*}

\begin{figure}
    \centering
    \includegraphics[width=0.8\linewidth]{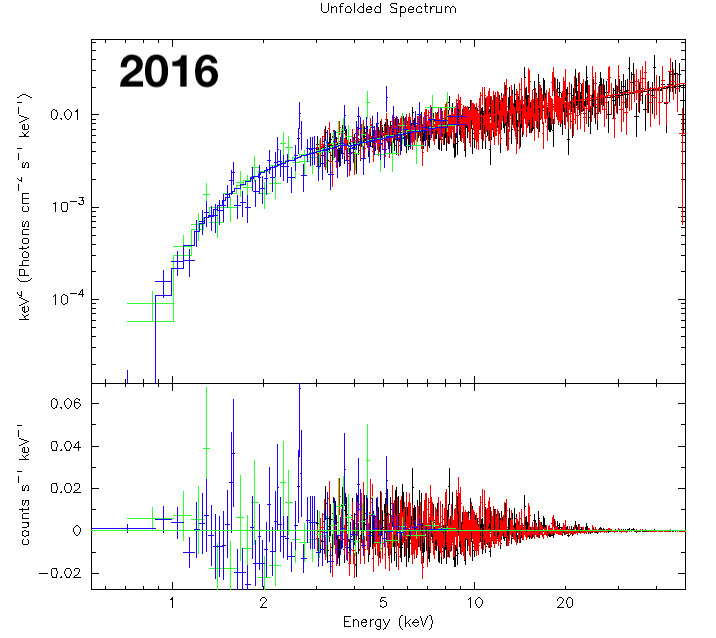}
    \includegraphics[width=0.8\linewidth]{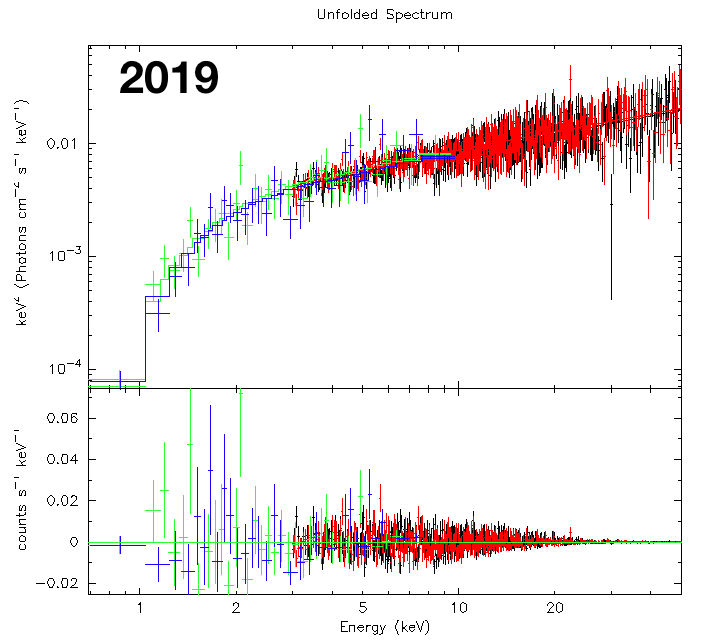}
    \caption{Spectral fit of \nustar FPMA and FPMB data (red and black) and the previous and subsequent (blue and green) \swiftxrt observation from 2016 (top) and 2019 (bottom). The spectrum is fitted with \texttt{constant$\,\times\,$ztbabs$\,\times\,$tbabs$\,\times\,$ cflux(powerlaw)} which is accounting for calibration differences between the instruments, assuming a power law for the source intrinsic spectrum, and absorption in the lensing galaxy as well as the Milky Way. Corresponding fit parameters are shown in Tab.\,\ref{tab:joint_2_spec}.}
    \label{fig:spec}
\end{figure}

We also modeled an absorbed log-parabola but did not find better fits than for the simpler and therefore preferable models described above. 
More sophisticated absorption models such as \texttt{tbnew}\footnote{\href{https://pulsar.sternwarte.uni-erlangen.de/wilms/research/tbabs/}{https://pulsar.sternwarte.uni-erlangen.de/wilms/research/tbabs/}} are not investigated because the focus of the current paper is on the continuum of \pks rather than absorption features.

By comparing the results we find that the \swiftxrt data alone poorly constrain the source properties and yield relatively large error bars for all fits, emphasizing that the \nustar data is crucial for studying this source. 
Considering the joint analysis, the absorbed power-law with N$_{\mathrm{H}}$ fixed to the Galactic value seems to be slightly disfavoured for the 2016 as well as the 2019 data (Tab.\,\ref{tab:joint_1_spec} bottom rows). 
The goodness of the fit is significantly improved by either leaving  N$_{\mathrm{H}}$ free to vary (Tab.\,\ref{tab:joint_1_spec} top rows) or adding a model component that accounts for absorption in the lensing galaxy (\texttt{ztbabs}, Tab.\,\ref{tab:joint_2_spec}). 
We acknowledge that the latter model includes more parameters but judge it to describe the most detailed and physically accurate model of the data. 
Thus, we utilize the parameters shown in Tab.\,\ref{tab:joint_2_spec} to perform the broad-band analysis of the intrinsic spectral energy distribution of \pks (see Sec. \,\ref{sec:sed}). 
The corresponding spectral fit of \texttt{constant$\,\times\,$ztbabs$\,\times\,$tbabs$\,\times\,$cflux(powerlaw)} to the joint X-ray data is shown in Fig.\,\ref{fig:spec} for the 2016 and 2019 data in the top and bottom, respectively.
The slightly different models determine similar values for the flux and the power-law index $\Gamma$, and our results indicate that there is very little change in the X-ray emission and spectral parameters of 2016 compared to the \gm-ray flaring phase of 2019.


\section{Analysis of the \nustar Time Series Data}
\label{sec:nustar_variab}
We analyzed the two \nustar observations of \pks,
introducing an unbinned method for detecting and characterizing variability. 
To our knowledge this approach has not been used previously.
Since the available measurements are relatively brief 
(22\,ks~$\approx$~6\,h for 2016 and 41\,ks~$\approx$~11\,h for 2019), 
any detected variability would be at a very short-timescale and probably
not relevant for the gravitational lensing -- but it 
is nevertheless valuable to search for it.
The Fourier transform is computed directly from the photon time-tags.
There is no need for time bins, and the usual but unnecessary 
requirement for ``bins large enough for a good statistical sample''
with its loss of information and time resolution.
The result can be evaluated at an arbitrary array of frequencies, 
which is not possible with standard implementations of the fast Fourier
transform (FFT). However, in most applications it is natural to evaluate 
it at the usual grid: integer multiples of the fundamental frequency 
(which is well defined) up to the Nyquist frequency \citep[which is not
well defined, c.f. ][]{VanderPlas_2018}.

\nustar has two identical focal plane modules, FPMA and FPMB,
which generate photon data streams that are statistically independent
(with respect to the random photon detection process).
This feature allows 
estimation of the power spectrum as the Fourier 
transform of one data stream times the complex conjugate of the Fourier
transform of the other data stream. The resulting "co-spectrum" was invented \citep{Bachetti_2018} to deal with deadtime effects,
which are unlikely to be a problem in these data.
Here we use the co-spectrum to provide another
estimate of the power spectrum of the source.

\subsection{Unbinned Fourier Analysis of Photon Data}
\label{sec:unbinned_ft}

The following formula \citep{Scargle_1989} yields the 
complex Fourier transform of sequential data 
with arbitrary sampling in time -- 
that is of a series of measurements 
$x_n(t_n)$ obtained  at any times $t_n$:
\begin{linenomath*}
    \begin{equation}
    \resizebox{0.93\linewidth}{!}{
        $F(\omega) = 
        \frac{
            \Sigma_{n} w_{n} x_{n} 
            \mbox{cos} [ \  \omega ( t_{n} - \tau(\omega) \ ]
        }{
            \{ \ { \Sigma_{n} w_{n} 
            \mbox{cos}^{2} [ \  \omega ( t_{n} - \tau(\omega) \ ]
            }\  \}^{1/2}
        }
        + i 
        \frac{
            \Sigma_{n} w_{n} x_{n} 
            \mbox{sin} [ \  \omega ( t_{n} - \tau(\omega) \ ]
        }{
            \{ \ { \Sigma_{n} w_{n} 
            \mbox{sin}^{2} [ \  \omega ( t_{n} - \tau(\omega) \ ]
            }\  \}^{1/2}
        }$
    } \ , 
    \label{ft_0}
    \end{equation}
\end{linenomath*}
\noindent
where
\begin{linenomath*}
    \begin{equation}
        \tau(\ \omega \ ) =
        \frac{1}{2 \omega}
        \mbox{arctan}\
        \frac{
            ( \ \Sigma_{n} 
            \ w_{n} \ \mbox{sin}\ 2  \omega t_{n} \ )
         }{
            ( \ \Sigma_{n} 
            \ w_{n} \ \mbox{cos}\ 2  \omega t_{n}\  )
         }
    \label{ft_tau}
    \end{equation}
\end{linenomath*}
\citep{Scargle_1989}\footnote{~Python implementation is available, for instance, at \href{https://github.com/astroml/gatspy}{https://github.com/astroml/gatspy}}. 
Here, $w_n$ is a set of statistical weights that can be applied to the individual data points. 

A straightforward extension to time-tagged photon data is obtained simply by writing $x_n(t_n) = \delta(t - t_n)$, that is delta functions at the photon detection times.
This is the same method adopted by
\cite{Yu_Peebles} to estimate 
the Schuster periodogram of point data.
Thus, we have

\begin{linenomath*}
    \begin{equation}
    \resizebox{0.93\linewidth}{!}{
        $F(\omega) = 
        \frac{ 
            \Sigma_{n} w_{n} 
            \mbox{cos} [ \  \omega ( t_{n} - \tau(\omega) \ ]
        }{
            \{ \ { \Sigma_{n} w_{n} 
            \mbox{cos}^{2} [ \  \omega ( t_{n} - \tau(\omega) \ ]
            } \ \}^{1/2}
        }
        + i 
        \frac{
            \Sigma_{n} w_{n} 
            \mbox{sin} [ \  \omega ( t_{n} - \tau(\omega) \ ]
        }{
            \{ \ { \Sigma_{n} w_{n} 
            \mbox{sin}^{2} [ \  \omega ( t_{n} - \tau(\omega) \ ]
            } \ \}^{1/2}
        }$
    } \ ,
    \label{ft_1}
    \end{equation}
\end{linenomath*}
\noindent
summed over the individual photon time-tags, and 
with a similar  form of  equation (\ref{ft_tau}) for the phase adjustment $\tau(\omega)$.

Power spectra can be immediately computed via 
\begin{linenomath*}
    \begin{equation} P(\omega) = F^{*}(\omega) F(\omega) =
        |  F(\omega)  |^{2}
        \label{lomb_scargle}
    \end{equation}
\end{linenomath*}
\noindent
where $F^{*}(\omega)$ is the complex conjugate of $ F(\omega)$. 
This expression is commonly known as the Lomb-Scargle periodogram \citep{Lomb_1976,Scargle_1982},
see also \citep{Barning_1963}, in wide use in astronomy and elsewhere.

Similarly the phase spectrum is 
\begin{linenomath*}
    \begin{equation} \phi(\omega) = arg( F(\omega) )
    \end{equation}
\end{linenomath*}
\noindent
and the co-spectrum \citep{Bachetti_2018} is
\begin{linenomath*}
    \begin{equation}
        \mbox{Co-spectrum}(\omega) = F_{A}^{*}(\omega) F_{B}(\omega)
    \label{cospec}
    \end{equation}
\end{linenomath*}
\noindent
in terms of the Fourier transforms of the data streams from the two
focal plane modules, A and B. 

\subsection{The NuSTAR \pks Data}

\begin{figure}[h]
\begin{center}
\includegraphics[width=0.8\linewidth]{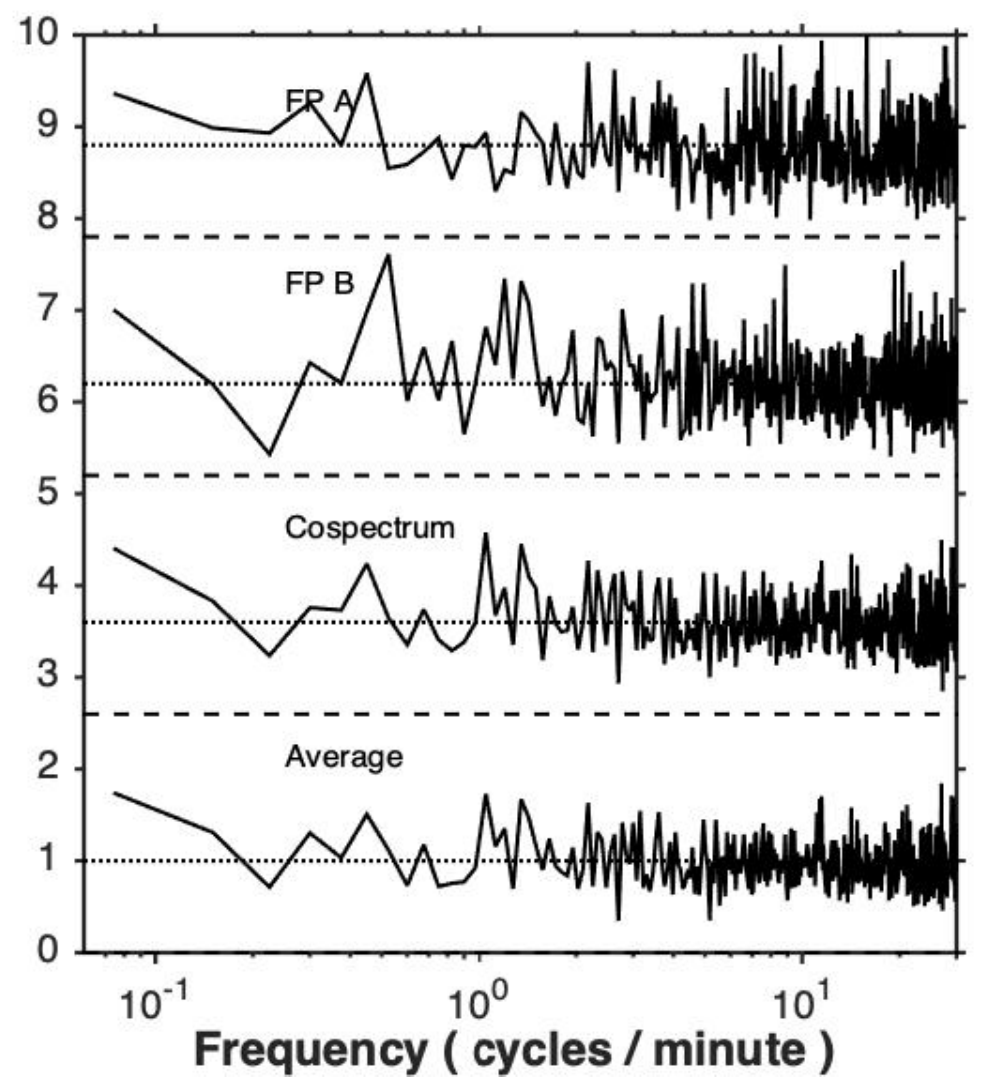}
\caption{Power spectra for one good-time interval (GTI) of the 2019 \nustar pointing.
The power spectra are from top to bottom:
FP A, FP B, the corresponding co-spectrum,
and the average of all three.}
\label{ps_obs_6}
\end{center}
\end{figure}

This procedure is applied to the \nustar data 
for the two observations 
described in Sect.\ref{sec:xray_spec}.
Shown in the Fig.\,\ref{ps_obs_6}
are the power spectra for the sixth
good-time-interval (GTI) of the 2019 data (ObsID: \textbf{8046}0628002, exposure $\approx$ 41\,ks),
since there was no evidence for 
variability over the course of the 
full observation interval.

Care must be taken in such averaging,
because the strong zero-frequency
peak (of order N, the
number of events) can
distort a simple average due to the 
variety of values of the fundamental 
frequency.
The approach used here to average
on a common, 
finely gridded frequency interval,
the spectra
evaluated at positive integer multiples
of the individual fundamental
frequencies but ignoring the
zero-frequency power in all cases.

Figure \ref{ps_obs_6} is a
linear plot of the power spectra,
on a log frequency scale.
The first two curves are for the 
two independent focal plane modules,
which should be identical except 
for photon-count statistics.
The third is the co-spectrum,
obtained as described above,
in Equation (\ref{cospec}).
The curve in the fourth panel 
is the average of the other three.
These curves have been offset vertically;
the horizontal dotted lines represent the 
value expected for white noise (unity)
and the dashed lines are the zero
of the power scale.

The results here are essentially the same
as for non-variable data, i.e. 
they are consistent with white noise.
As a \textit{null case} we applied the algorithm specified in Eq.\,(\ref{ft_1}) to \nustar time-tagged photon data of the Perseus galaxy cluster, a constant source on the relevant timescales (see Sec.\,\ref{sec:data_nustar}). 
Also in this case, the resulting power spectra show only statistical fluctuations consistent with white noise\footnote{To an extent, this verifies the claim in the \nustar data analysis documentation that the timing `` ... is kept stable to within a few milliseconds by correcting the drift of the spacecraft clock relative to UT clocks at the ground stations with additional corrections performed at the SOC.'' See \nustar Guest Observer Program, 
\nustar Observatory Guide, Section 3.6,  
\href{https://heasarc.gsfc.nasa.gov/docs/nustar/NuSTAR_observatory_guide-v1.0.pdf}{https://heasarc.gsfc.nasa.gov/docs/nustar/NuSTAR\textunderscore observatory\textunderscore guide-v1.0.pdf}} 
similar to Fig.\,\ref{ps_obs_6}.

The conclusion is that \pks 
does not show significant variation 
over the time scales corresponding 
to the \nustar observations, namely
from hours to milliseconds.

\noindent


\section{Delay in the \gm-ray light curve of \pks}
\label{sec:lag}

As pointed out in the Section \ref{sec:intro_1830},
it is not possible to resolve
the two images at \gm-ray energies,  
so we are working with 
a superposition of two 
light curves -- taken to be 
identically shaped, 
except for the observational errors, 
but offset from each other
in time and amplitude.
To determine these lens observables, 
we pursue two approaches:
the auto-correlation function
with an improved analysis procedure
and a modeling approach using 
a novel exact solution for $x(t)$
in Equation~(\ref{eq:lc_model}).
We use the full observation interval as well as individual segments of the light curve (see Sec.\,\ref{sec:data_fermi}) to study potential changes of these parameters over time.

We also investigated a method based on individual flares similar to the "maximum peak method" \citep{Barnacka_2011, Barnacka_2015}.
Our approach was to define flares based on the Bayesian block and HOP algorithm \citep{Wagner_2021} and check whether there was an accumulation of peak separations at a certain lag.
However, this method relied on a limited number of prominent flares and was thus more sensitive to noise and irregular variability (for example induced by microlensing) than the methods employing the whole light curve. 
The approach led to inconclusive results \citep[see][for discussion]{Wagner_phd}.

Appendix \,\ref{sec:MO},
describes another 
parameter estimation 
method,
based on optimizing a formal
solution of Equation (1)
we feel is 
worthy of further development.

\subsection{The Autocorrelation Method}
\label{sec:acf_method}

\subsubsection{The Autocorrelation Function (ACF)}
The basic delay Equation 
(\ref{eq:lc_model}) 
yields the following
relation between the
autocorrelation functions of the 
observed and intrinsic light curves,
$R_{Y}$ and $R_{X}$ respectively:
\begin{linenomath*}
\begin{equation}
R_{Y}(\tau) = 
( 1 + a^{2} )  R_{X}(\tau )
+
a [ R_{X}(\tau + t_{0} ) +  R_{X}(\tau - t_{0} ) ] 
\label{eq:acf_model}
\end{equation}
\end{linenomath*}
\noindent
(see Eq.\,\ref{eq:acf_model_append} 
for the derivation).
This relation exhibits an interesting property:
\textit{the (two-sided) autocorrelation function of the observed light curve 
is the sum of three scaled and shifted versions of 
the autocorrelation function of the 
source light curve.}

The first term, peaking at zero lag, characterizes 
short- and long-term memory effects in the source -- 
as if it were unlensed.
The other two terms, 
peaking at the lagged values $\pm t_{0}$,
combine to make a symmetric contribution to $R_{Y}(\tau)$
that provides information about the delay.
This formula 
can be considered as 
justification for the common approach 
in which a peak in the ACF is
taken as evidence for a delay at the corresponding lag.
We are unaware of any previous justification for 
such ``bump hunting'' with
autocorrelations.

To compute the ACF, we take the inverse Fourier transform of the power spectrum obtained with the Lomb-Scargle Periodogram (LSP, see Sec.\,\ref{sec:unbinned_ft}).
The resulting ACF is shown for small lags ($\tau<100$) as black solid line in the top panel of Figure \ref{fig:envelope}.
A detailed discussion on other computation methods for unevenly sampled time series is presented in Appendix\,\ref{sec:gaps}.

\subsubsection{The Lower Envelope (Excess)}
In searching for potentially weak 
peaks superimposed on the ACF, 
it is useful to establish a baseline curve 
representing the broad sweep of the main ACF.
To avoid as much as possible 
the subjectivity of 
arbitrary choices that standard 
detrending methods introduce,
we adopted a parameter-free method:
the upper convex hull\footnote{~The smallest convex polygon that contains all the given points, formed by connecting the outermost points while maintaining convexity, \href{https://docs.scipy.org/doc/scipy/reference/generated/scipy.spatial.ConvexHull.html}{scipy.spatial.ConvexHull}.}, which is  the ``outline'' of the set of points, formed by connecting the outermost points in a way that no part of the line goes inside the points.
We refer to this quantity as the lower envelope (LE) throughout this paper.

Like most AGN, \pks
has a power spectrum that is 
approximately a power law.
While exact results for the ACF of a 
process that has such a $f^{-\alpha}$
power spectrum are not known, 
in many cases 
a power law ACF is a good approximation
\citep{Carpena_2022}.
If the intrinsic ACF of \pks 
were an exact power law, 
its lower envelope would 
in fact be the same as 
the true ACF itself
(this is true for any monotonic ACF).
Lensing delays would then yield 
an excess to the envelope,
as quantified in Eq. (\ref{eq:acf_model}).

\begin{figure}
    \centering
    \includegraphics[width=0.9\linewidth]{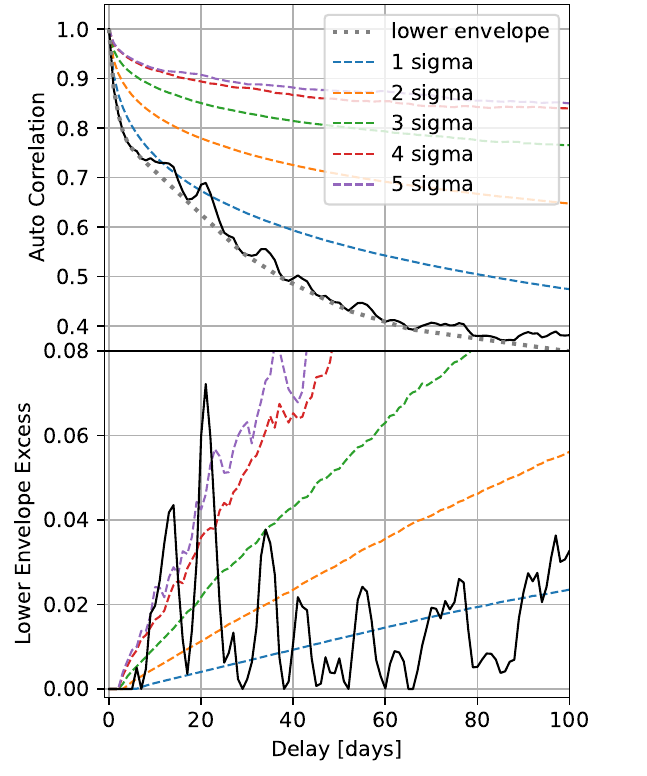}
    \caption{Top: Autocorrelation
    of daily binned, filtered \pks \fermilat data (black solid line) 
    and its lower envelope (LE, gray dotted line).
    Bottom: the lower envelope excess (LEE, black solid line).
    Visible in both panels,  the largest peak at 21 days is presumed to be due to strong lensing.
    The other peaks are probably signatures of correlated variability of \pks itself, see Sec.\,\ref{sec:discuss}.
    Significance lines were derived with percentiles, see text.
    }
    \label{fig:envelope}
\end{figure}

In reality, even if the 
underlying stochastic astrophysical process is 
truly 
$f^{-\alpha}$ noise, 
any one realization of it (e.g.
the observed light curve of interest here)
will contain fluctuations that modify
this picture.
Hence the 
lower envelope of the ACF
is an attractive way to define such a baseline,
above which the shifted,
positive-definite excess will protrude. 
Figure \ref{fig:envelope} depicts the lower envelope as gray dotted line in the top panel and the corresponding lower envelope excess (LEE = subtraction of LE from the ACF) as solid black line in the lower panel. 
The LEE is used throughout  the
analysis below, for two reasons:
it enhances 
any excess above the smooth baseline
of the lower envelope;
it also eliminates 
the sloping background, 
which, unless corrected for, 
could yield a small bias in ACF peak locations.

\subsubsection{Significance of Delay}
To asses the significance of auto-correlation at a given lag, we repeat the procedure for 50,000 simulated mock light curves following \cite{TimmerKoenig_1995}.
Based on the LSP of the original data, we estimate a PSD slope of $-1.27$ and generate continuous colored noise with the same exponent\footnote{~Colored noise generates simulated data by applying a power law exponent to the amplitude of each frequency component, assigning random phases to each frequency to ensure a real-valued output, and then performing an inverse Fourier transform to reconstruct the time-domain signal with the desired spectral slope, \href{https://github.com/felixpatzelt/colorednoise}{colorednoise}}.
For each ACF bin, we assess the significance based on the percentiles, with 1/2/3/4 sigma corresponding to the 68th/95th/99.7th/99.9th percentile, respectively, as shown in the top panel of Fig.\,\ref{fig:envelope}.
This procedure has been applied in many studies in this context, but it is important to emphasize that the derived significance lines depend sensitively on the imposed properties of the simulated data, i.e. the PSD slope.
Moreover, the direct application only addresses the auto-correlation broadly and typically underestimates the significance of a relatively narrow feature.
What we are truly looking for, is an excess of correlation on top of the intrinsic correlated noise behavior, which we approximate with the lower envelope.
We therefore repeat the procedure of generating these significance lines based on the lower envelope subtracted from the ACF, i.e. the lower envelope excess (LEE), which is shown in the bottom panel of Fig.\,\ref{fig:envelope}.
This illustrates that while the absolute value of the simulated auto-correlations can be overall high, the fluctuation generating the peaks around 13 and 21 days are highly significant ($>5\sigma$).
In this work, we focus on one main delay which is expected for two unresolved images.
We discuss potential causes for the smaller significant delay in Sec.\,\ref{sec:discuss}.

\subsubsection{Deriving the lens observables}
The first lens observable, the delay $t_0$, is assessed through a parabolic fit of the most prominent peak in the LEE.
The obtained results for the full light curve as well as the sub-intervals are reported in Tab.\,\ref{tab:acf_magnification} of Sec.\,\ref{sec:lag_results}.

In Appendix\,\ref{sec:extract_a}, we present a novel approach to also extract the second lens observable, the magnification factor $a$, from the ACF.
This is based on its expected shape characterized with Eq.\,\ref{eq:acf_model}.
The principle is to evaluate this expression at lags $\tau=0$ and $\tau=t_{0}$ and approximate values of $R_{X}(\tau)$ and $R_{Y}(\tau)$ from the ACF and our LE in Fig.\,\ref{fig:envelope}.
In doing so, we consider the effect of observational errors on the ACF, as illustrated in Fig.\,\ref{fig:acf_spike}.
The procedure yields two solutions for $a$ (Eq.\,\ref{eq:aa_acf_solution})
which are reciprocals of one another, as required by the time-reversal invariance.
The magnification ratios approximated with this approach for the whole light curve and its sub-intervals are shown in Tab.\,\ref{tab:acf_magnification} of Sec.\,\ref{sec:lag_results}.

\subsection{Results: Estimated values of $a$ and $t_{0}$ over time}
\label{sec:lag_results}
\begin{figure}[t]
\centering
\includegraphics[width=0.99\linewidth]{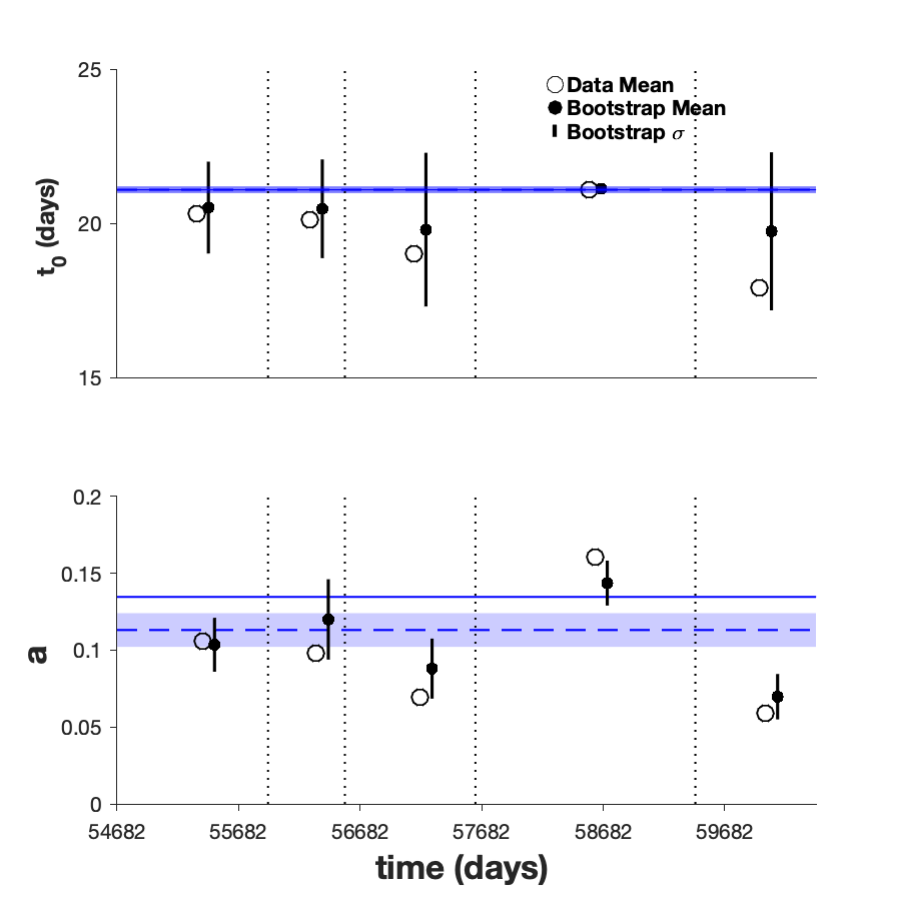}
\caption{
Estimastes for the two \pks lens observables plotted as circles along with the corresponding bootstrap means and 
standard deviations plotted as filled circles and error bars. 
Top: the delay $t_{0}$ from polynomial fits of the LEE; bottom: magnification ratio $a$ obtained using Eq.\,(\ref{eq:aa_acf_solution}).
These points 
are for 5 subintervals (edges at MJD 55880, 56477, 57828, and 59668 indicated by vertical dotted lines).
Horizonatal blue lines are for the whole 
observation interval:
mean from the data (solid),
bootstrap mean (dashed),
and bootstrap standard deviation
(filled bar);
these are nearly invisible 
for the delay values.
} 
\label{fig:results_1830}
\end{figure}

\begin{table}[]
\centering
\begin{tabular}{c | c c c | c c c}
\hline
\multicolumn{7}{c}{\textbf{ACF ANALYSIS: LAGS \& MAGNIFICATION RATIOS}} \\
\hline
\textbf{start-end} & $\mathbf{t_0}$ & \multicolumn{2}{c|}{\textbf{Bootstrap}} & $\mathbf{a}$ & \multicolumn{2}{c}{\textbf{Bootstrap}} \\
 & & Mean & Std & & Mean & Std \\
\hline
54684-55880 &  20.30 &  20.73 &   1.63 &  0.11 &  0.10 &  0.02 \\ 
55880-56477 &  20.10 &  20.37 &   2.21 &  0.10 &  0.11 &  0.02 \\ 
56477-57828 &  19.00 &  20.24 &   2.56 &  0.07 &  0.09 &  0.02 \\ 
57828-59668 &  21.10 &  21.22 &   0.14 &  0.16 &  0.14 &  0.01 \\ 
59668-60433 &  17.90 &  19.78 &   2.70 &  0.06 &  0.07 &  0.02 \\ 
54684-60433 &  21.10 &  21.12 &   0.12 &  0.13 &  0.12 &  0.01 \\ 
\hline
55422-55622 &  20.10 &  20.50 &   0.78 &  0.18 &  0.18 &  0.03 \\ 
56052-56202 &  20.20 &  20.57 &   1.65 &  0.13 &  0.16 &  0.04 \\ 
56802-57002 &  16.00 &  19.83 &   3.14 &  0.09 &  0.17 &  0.05 \\ 
57012-57102 &  19.10 &  19.88 &   1.67 &  0.22 &  0.23 &  0.07 \\ 
\hline
\end{tabular}
\caption{Auto-correlation Analysis: Lags and Magnification Ratios for this works sub-intervals (row 1-5), the whole light curve (row 6), and the intervals selected in \cite{Barnacka_2015} for comparison. We derive the lag $t_0$ and magnification ratio $a$ as described in Sec.\,\ref{sec:acf_method} and obtain a bootstrap mean and standard deviation based on Appendix\,\ref{sec:bootstrap}.}
\label{tab:acf_magnification}
\end{table}

The autocorrelation  
method was applied to the full \fermilat light curve of \pks,
as well as the five sub-intervals (separating the light curve as illustrated in Fig.\,\ref{fig:lcs})
to assess possible variation 
of the lens observables.
The error estimates are 
based on 1000 random 
re-samplings with replacement, 
in the usual bootstrap fashion as described in Appendix\,\ref{sec:bootstrap}.

The complete coverage means 
that we are using all of the
information presented by the
data, and 
the lack of overlap 
means that the data in these
sub-intervals are
\textit{statistically independent 
of each other 
with respect to observational
errors}.
Therefore possible variation
in the lensing observables
over time can be assessed 
using the 
statistical errors in 
the parameter estimates 
in the sub-intervals.\footnote{The fluxes are not statistically independent
of each other with 
respect to the 
random variability of the source.
That is to say, 
on various time scales
the intrinsic variation 
of \pks contains 
causal physical effects
that yield autocorrelations
or ``memory.''
Failure to distinguish 
these two stochastic processes -- 
(a) source flux variations and
(b) observational errors -- 
sometimes leads to confusion.
Here the point is that 
the statistical distribution 
of our parameter estimates 
derives only from (b),
and are thus independent 
from one sub-interval to another.
In other words, 
stochastic process (a) 
is the one that is under 
study here,
as limited by the uncertainty
yielded by observational errors, process (b).}

The results of the ACF procedure
are displayed in Table\,\ref{tab:acf_magnification}.
The first 5 lines of the table cover the 
sub-intervals,
whereas line 6 is for the full observation period.
For comparison, we apply our analysis to the 
intervals studied by 
\cite{Barnacka_2015} 
and list the results in 
the last four lines of the table.

The first two columns 
give the times at the beginning and end 
of the sub-intervals, in MJD.
The next three columns display the delay estimate,
followed by its bootstrap mean and
standard deviation.
The final three columns give the same 
values for the magnification ratio.

The bootstrap standard deviations
are $1\sigma$ uncertainties;
they are relatively small.
The strong flaring in interval 4 
apparently dominates the lag determination,
yielding very similar values for the lensing
observables in this sub-interval 
and in the 
whole interval.

We plot the obtained values for the ACF method and its bootstrap mean as hollow circles and squares, respectively, in Figure~
\ref{fig:results_1830}.
The bootstrap variances described 
in Appendix \ref{sec:bootstrap} 
can be considered error bars and 
are plotted as vertical, solid, black lines.

There is no compelling 
indication of 
time variability of the lag,
considering the error bars.


\section{Broad-band spectral energy distribution}

\label{sec:sed}

\begin{figure*}
\includegraphics[width=\textwidth]{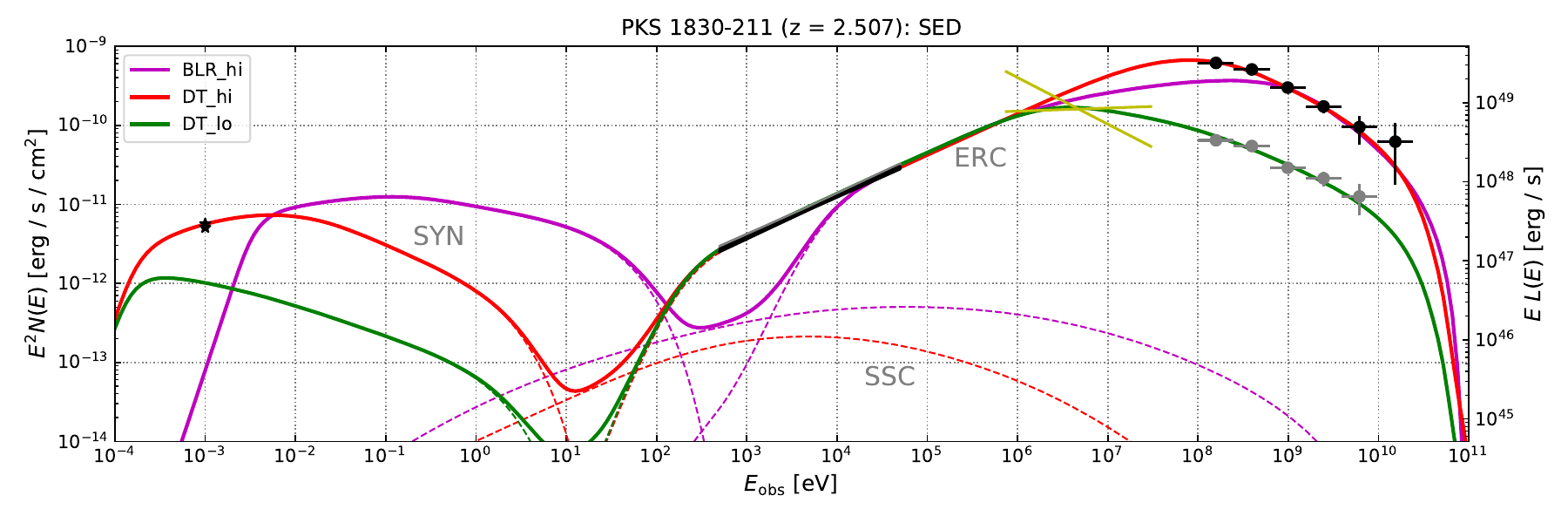}
\caption{Spectral energy distribution of \pks. Observational data are indicated for the 2019 March epoch (black symbols), the 2016 April epoch (gray symbols), and for other epochs (yellow symbols).}
\label{fig:sed}
\end{figure*}

Figure \ref{fig:sed} presents the SED of \pks.
For the flaring state in 2019 March (black symbols), we present a HE \gm-ray spectrum from the \fermilat, an X-ray power-law model fitted jointly to the \nustar and \swiftxrt data, and millimeter flux density resolved by ALMA into two images \citep{Marti-Vidal_2020}. 
The latter was observed on 2019 April 11, which was one month after the 2019 \nustar observation. 
For the quiescent state in 2016 (gray symbols), we present a HE \gm-ray spectrum from \fermilat and an X-ray power-law model which largely overlaps with the model for 2019 data.
We also indicate archival $\sim$MeV \gm-ray data from Compton/COMPTEL \citep{Zhang_2008} with yellow symbols.

The observed SEDs are modelled with non-thermal leptonic emission calculated with the code {\tt BLAZAR} \citep{Moderski_2003}.
The emitting region is assumed to have the shape of a sphere expanding along a conical jet with uniform bulk Lorentz factor $\Gamma_{\rm j}$ and half-opening angle $\theta_{\rm j}$.
As the shell propagates a radial distance from $r_0/2$ to $r_0$, energetic particles are injected with a broken power-law energy distribution with $N(\gamma) \propto \gamma^{-p_1}$ for $\gamma_{\rm min} < \gamma < \gamma_{\rm br}$ and $N(\gamma) \propto \gamma^{-p_2}$ for $\gamma_{\rm br} < \gamma < \gamma_{
\rm max}$.
These particles undergo radiative and adiabatic cooling, and radiation signals are integrated over the entire injection distance towards an observer located at viewing angle $\theta_{\rm obs}$.

The SED model consists of three main components: synchrotron (SYN), synchrotron self-Compton (SSC) and external radiation Compton (ERC).
External radiation fields include two main components: broad emission lines (BEL) produced in the broad line region (BLR), and thermal infrared emission (IR) produced in the dusty torus (DT).
The radial profiles of energy densities $u_{\rm BEL}(r)$ and $u_{\rm IR}(r)$ follow the standard model described in \cite{Sikora_2009}.

\pks is a very luminous blazar belonging to the class of FSRQs.  Its apparent $\nu L_\nu$ luminosity reaches $\simeq 3\times 10^{49}\;{\rm erg\,s^{-1}}$ at $100\;{\rm MeV}$ in the 2019 state (which according to the \gm-ray light curve is not exactly the peak of the flare), but also $\simeq 3\times 10^{48}\;{\rm erg\,s^{-1}}$ in the 2016 state, and $\simeq 1.5\times 10^{48}\;{\rm erg\,s^{-1}}$ in the hard X-rays.
Such high apparent luminosities could be boosted not only due to relativistic motion along the jet --- we adopted a fairly high bulk Lorentz factor of the jet $\Gamma_{\rm j} = 30$ (cf. $\Gamma_{\rm j} = 10-15$ in \citealt{Donnarumma_2011}, $\Gamma_{\rm j} = 20$ in \citealt{Abdo_2015}, $\Gamma_{\rm j} = 35$ in \citealt{Abhir_2021}, $\Gamma_{\rm j} = 18$ in \citealt{Vercellone_2024}) --- but also due to strong gravitational lensing.
Following \cite{Donnarumma_2011}, \cite{Abdo_2015} and \cite{Neronov_2015}, we assume a fixed magnification factor of $\mu = 10$, by which the SEDs calculated by the {\tt BLAZAR} code are multiplied.
This is a moderate value when compared with individual strongly lensed supernovae, for which total lensing magnifications up to $\sim 50$ have been inferred \citep[e.g.,][]{Goobar_2017}.

In a scenario typical for FSRQs, we attempt to connect the observed X-ray and \gm-ray spectra with a single ERC component. This is strongly motivated by the observed spectral stability of the X-ray emission.  In particular, the two \nustar observations returned almost the same hard X-ray SEDs (including the same normalizations), despite order-of-magnitude variations in the \gm-ray flux.  This is incompatible with a significant contribution of the SSC component to the soft X-ray band \citep{Abdo_2015}, and we therefore constrained the SSC component to be at least one order of magnitude below the observed soft X-ray flux density.

\begin{figure}
\centering
\includegraphics[width=\columnwidth]{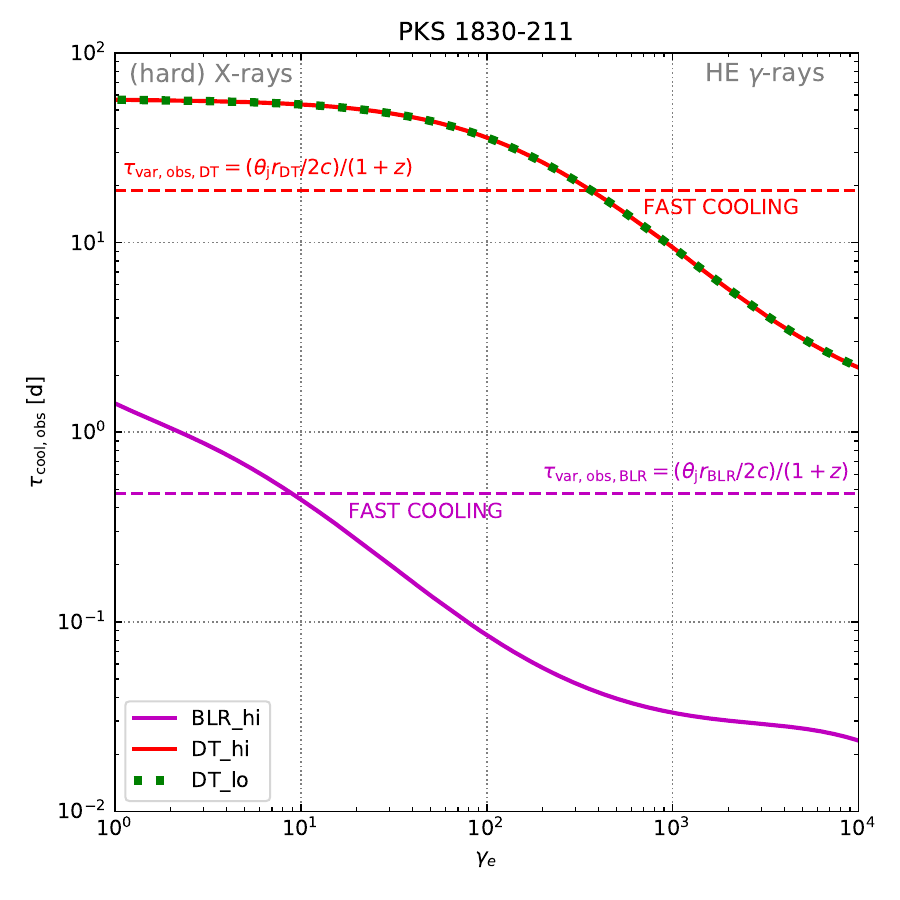}
\caption{Apparent cooling time scales $\tau_{\rm cool,obs}$ as functions of electron energy $\gamma_{\rm e} m_{\rm e}c^2$ for the SED models presented in Figure \ref{fig:sed}.  Low-energy electrons with $\gamma_{\rm e} \sim 1-10$ produce (hard) X-ray emission via ERC mechanism; high-energy electrons with $\gamma_{\rm e} \sim 10^3 - 10^4$ produce HE \gm-ray emission via ERC.  For comparison, apparent variability time scales $\tau_{\rm var,obs}$, corresponding roughly to light travel time across the jet at characteristic radii $r_{\rm BLR}, r_{\rm DT}$, are indicated with the dashed lines.}
\label{fig:tcool}
\end{figure}

We considered two characteristic distance scales $r_0$ for the location of the emitting region along the jet.
In model BLR\_hi, we adopted the BLR radius \citep{Ghisellini_2008, Sikora_2009} $r_0 = r_{\rm BLR} \simeq 0.1 L_{\rm d,46}^{1/2}\;{\rm pc} \sim 0.084\;{\rm pc}$,
where $L_{\rm d,46} \equiv L_{\rm d}/(10^{46}\;{\rm erg\,s^{-1}}) \sim 0.7$ is the accretion disk luminosity \citep{Celotti_2008, Abhir_2021} (cf. $L_{\rm d,46} \sim 0.4$ in \citealt{Abdo_2015}),
at which the external radiation fields are dominated by broad emission lines of luminosity $L_{\rm BEL} \sim 0.1 L_{\rm d}$, energy density $u_{\rm BEL} \simeq L_{\rm BEL}/(4\pi cr_{\rm BLR}^2) \sim 0.028\;{\rm erg\,cm^{-3}}$, and characteristic photon energy $E_{\rm BEL} \sim 10\;{\rm eV}$.
For $\gamma_{\rm min} = 1$, the ERC(BEL) component extends from the observed photon energy $E_{\rm ERC(BEL),min} \sim (2\Gamma_{\rm j} \gamma_{\rm min})^2 E_{\rm BEL}/(1+z) \sim 10\;{\rm keV}$.
Hence, when matched to the hard X-ray spectrum measured by \nustar, this component cannot reproduce the observed soft X-ray emission.
In the high-energy \gm-rays, this component can be matched to the \fermilat spectrum above $\sim 1\;{\rm GeV}$, but it underproduces the $\sim 100\;{\rm MeV}$ flux density due to an onset of a cooling break at $\sim 1\;{\rm MeV}$ (in addition to a break at $\sim 0.8\;{\rm GeV}$ obtained from the break in the injected electron energy distribution at $\gamma_{\rm br} = 600$ from power-law index $p_1 = 1.85$ to $p_2 = 3.2$).
Confirmation that this is a cooling break can be found in Figure \ref{fig:tcool}, which shows the radiative cooling time scale converted into the observer's frame $\tau_{\rm cool,obs}$ as a function of electron energy $\gamma_{\rm e} m_{\rm e}c^2$.
The cooling time scale can be compared with the characteristic variability time scale $\tau_{\rm var,obs}$ corresponding to the light travel time across the jet at distance scale $r_{\rm BLR}$.
Electrons with $\gamma_{\rm e} > 10$ have $\tau_{\rm cool,obs} < \tau_{\rm var,obs}$, hence they are in the fast cooling regime, the onset of which results in a break of the electron distribution function at $\gamma_{\rm cool,br} \sim 10$.

In Model DT\_hi, we adopted the DT radius \citep{Nenkova_2008a, Sikora_2009} $r_0 = r_{\rm DT} \simeq 4 L_{\rm d,46}^{1/2}\;{\rm pc} \sim 3.3\;{\rm pc}$, at which the external radiation fields are dominated by thermal infrared ($E_{\rm IR} \sim 0.3\;{\rm eV}$) emission from the dusty torus.
We adopt the torus luminosity of $L_{\rm IR} \sim 0.1L_{\rm d}$, which corresponds to an energy density $u_{\rm IR} \simeq L_{\rm IR}/(4\pi cr_{\rm IR}^2) \sim 1.7\times 10^{-5}\;{\rm erg\,cm^{-3}}$.
The ERC(IR) component extends from the observed photon energy $E_{\rm ERC(IR),min} \sim 0.3\;{\rm keV}$, hence it can readily explain both the soft and hard X-ray emission as a single power-law.
Furthermore, because of much lower energy density of external photons, the cooling break increases to $\gamma_{\rm br,cool} \sim 350$, not far below the cooling break of $\gamma_{\rm br} = 1200$ from $p_1 = 1.95$ to $p_2 = 3.3$ imposed on the injected electron energy distribution in order to match the HE \gm-ray spectrum observed with \fermilat.

We also present Model DT\_lo located at $r_{\rm DT}$ and matched to the low \gm-ray state observed in 2016.
The main difference from Model DT\_hi is a much lower break in the electron distribution $\gamma_{\rm br} = 130$, which causes the ERC(IR) component of the SED to peak in the 1-10 MeV range, where it matches closely the historic result from Compton/COMPTEL.

Table \ref{tab:sed_model_params} presents components of the minimum jet power implied by our SED models, estimated as $L_{\rm X} = \pi(r\theta_{\rm j})^2 \Gamma_{\rm j}^2 u_{\rm X}'c$, where $u_{\rm X}'$ is the corresponding co-moving energy density.
The contribution of magnetic field is estimated from its co-moving strength.
The contributions of electrons and radiation are estimated by integrating their energy distributions.
The contribution of protons is estimated under the assumption of one cold proton for every electron \citep{Ghisellini_2014}, alternatively it can be reduced by allowing for copious electron-positron pairs \citep{Pjanka_2017}.
In all models we find strong domination of protons, exceeding by factor $\sim 200 - 300$ the contribution of electrons, and reaching values above $10^{48}\;{\rm erg\,s^{-1}}$ for the DT models.
Assuming the mass of the AGN black hole of $M_{\rm BH} \sim 10^9 M_\odot$, the Eddington luminosity of $L_{\rm Edd} \sim 1.6\times 10^{47}\;{\rm erg\,s^{-1}}$ would be strongly exceeded by the proton jet power.
In order to reduce the proton jet power below $L_{\rm Edd}$, especially the DT models would require a pair content equivalent to $n_{\rm e}/n_{\rm p} \sim L_{\rm p}/L_{\rm Edd} \gtrsim 30$.
Note that the electron distribution is strongly dominated by low-energy particles with average Lorentz factors $\left<\gamma\right>_{\rm inj} \sim 10$ for the injected distribution, and $\left<\gamma\right> \sim 7$ for the evolved distribution.
The contribution of magnetic fields is roughly at the level of equipartition with electrons in the BLR model, $L_{\rm B}/L_{\rm e} \sim 0.5$, and strongly sub-equipartition in the DT models, $L_{\rm B}/L_{\rm e} \sim 0.003$.
The radiative efficiency of the BLR model is high, with $L_{\rm r} \sim L_{\rm e}$, consistent with fast cooling; while the efficiencies of the DT models are much lower, with $L_{\rm r}/L_{\rm e} \sim 0.1$, consistent with slow cooling.

\begin{table}
\caption{Selected parameters of the SED models presented in Figure \ref{fig:sed} and discussed in the text. Other important parameters common to all models are: lensing magnification $\mu = 10$, jet bulk Lorentz factor $\Gamma_{\rm j} = 30$, jet half-opening and viewing angles $\theta_{\rm j} = \theta_{\rm obs} = 1/\Gamma_{\rm j}$, minimum Lorentz factor of electrons $\gamma_{\rm min} = 1$.}
\label{tab:sed_model_params}
\centering
\begin{tabular}{cccc}
\hline\hline
Model & BLR\_hi & DT\_hi & DT\_lo \\
\hline
$r\;{\rm[pc]}$ & 0.084 & 3.34 & 3.34 \\
$B'\;{\rm[G]}$ & 2.25 & 0.044 & 0.037 \\
$\gamma_{\rm br}$ & 600 & 1200 & 130 \\
$\gamma_{\rm max}\;[10^4]$ & 1.0 & 1.5 & 1.5 \\
$p_1$ & 1.85 & 1.95 & 1.95 \\
$p_2$ & 3.2 & 3.3 & 3.1 \\
\hline
$\log_{10}(L_{\rm e}\;{\rm[erg/s]})$ & 45.3 & 46.3 & 46.2 \\
$\log_{10}(L_{\rm p}\;{\rm[erg/s]})$ & 47.9 & 48.7 & 48.7 \\
$\log_{10}(L_{\rm B}\;{\rm[erg/s]})$ & 45.1 & 44.9 & 44.7 \\
$\log_{10}(L_{\rm r}\;{\rm[erg/s]})$ & 45.3 & 45.5 & 44.9 \\
\hline\hline
\end{tabular}
\end{table}

\section{Discussion}
\label{sec:discuss}

In the radio regime \pks is detected to show two gravitationally-lensed, delayed images, which are about 1 arc sec apart and therefore not resolved by \fermilat.

In principle the gravitational lensing is 
achromatic, so any departure from exact agreement of the time 
delay measured in various bands would imply discrepant locations and/or 
sizes of regions responsible for emission in various bands.  
In particular, regarding the \gm-ray data, an absence of a 
signature of such delay in a highly variable source would have a 
profound implications on the structure of the \gm-ray emitting region,
as pointed out by \cite{Barnacka_2015}. Specifically, those authors 
argue that if the time delay observed at different epochs is not 
constant, this might be caused by the \gm-ray emission taking 
place at very different regions of the jet, and specifically, at 
kiloparsec distances. Thus, the determination of the delay is crucial in understanding the jet's nature and location of the dissipation region of its kinetic energy.

We analyzed the 16 year \fermilat light curve of \pks to search for imprints of the expected time delay.
We improve the ACF approach to determine lags with subtraction of the LE (Sec.\,\ref{sec:acf_method}).
This highlights the truly relevant excess of potential delays on top of the broad sweep ACF, which is expected for colored noise.
From this we obtain for the most prominent delay $t_0=21.1 \pm 0.1\,$d at $>5\sigma$ confidence and in a novel extraction procedure a magnification factor $a=0.13 \pm 0.01$.

In a forthcoming work, \cite{Buson_inprep} separate the daily binned Fermi-LAT light curve of PKS in six independent segments of fewer than 200 days and analyze the delay using the discrete autocorrelation method, after removing trends on a 15-day timescale in the time domain. This approach is complementary to our procedure and yields consistent results.

The ACF method also yields a significant delay at 15\,d.
We can not rule out that this could be cause by a potential third image, which could be scrutinized in future work.
However, radio observations show a third image which is 140 times dimmer than the two main images, which are comparably bright, so it seems unlikely that a third image would show such drastic effects in the \gm-rays.
As mentioned above, micro-lensing would typically only affect one image and can thus not explain a consistent 15\,d delay.
However, \cite{Tie_2018} show that micro-lensing can cause the delay to vary but this contradicts that we find a consistent main delay and the expected timescales for this would be larger \citep[i.e. from several months to several years][]{Millon_2020}.
We deem most likely that \pks itself shows auto-correlated behavior. This could be by chance due to random, intrinsic flaring or rooted in physical properties such as a binary system of super-massive black holes.
Verification of this exceeds the scope of this paper and is left for future work.

The constancy of 
the delay throughout the 16-year \textit{Fermi} observations (see Fig.\,\ref{fig:results_1830})
implies a roughly constant location of the emission region - and allows 
us to model the broad-band spectral energy 
distribution of both observations of the object.  Since the object 
is lensed into multiple images, its apparent flux is magnified, and, 
following \cite{Neronov_2015} we adopt the magnification factor of the 
summed images of 10, reasonably common value for objects which are 
multiply-imaged by an intervening galaxy. We considered the 
analysis of two broad-band spectra of the object, both coincident 
with two \nustar pointings, in 2016 and 2019, when the \gm-ray 
flux of the source was at very different levels (increased 
by a factor of 10).  We note that the \nustar spectrum was crucial in 
determining the continuum, mainly because the spectral analysis of the 
X-ray data at lower energies suffered from heavy absorption by both 
our Galaxy as \pks is close to the Galactic plane, and 
by the lensing galaxy. We apply a model including 
the synchrotron radiation for the low-energy component, and inverse Compton 
model for the high-energy component, based on the {\tt blazar} code 
\citep{Moderski_2003}. Our analysis implies that the model most consistent 
with our data includes inverse Compton emission by photons from the dusty 
torus up-scattered by the same electron population which are producing the 
synchrotron emission, while the synchrotron self-Compton emission is significantly sub-dominant.  

A recent work by \cite{Vercellone_2024} investigated the same \gm-ray flare of PKS 1830-211 using additional \gm-ray data from AGILE.
They modeled the flaring SED with a two-zone leptonic model, in which the X-ray and \gm-ray parts are fitted with different components (Comptonization of infrared emission from the dusty torus and broad emission lines, respectively).
The key differences from our models are: (1) a significantly lower bulk Lorentz factor $\Gamma_{\rm j} = 18$; (2) locating the emitting region deeper within the BLR, at distance scale $0.03\,{\rm pc}$; (3) a weaker magnetic field $B' \sim (0.2 - 1)\,{\rm G}$ at such distance; (4) a narrow energy range of electrons with $\gamma_{\rm min} = 30$ and $\gamma_{\rm max} \sim (1-3)\times 10^3$.
This results in much lower requirements for the powers of protons $\log_{10}(L_{\rm p}\;{\rm[erg/s]}) \sim 46.3$ and electrons $\log_{10}(L_{\rm e}\;{\rm[erg/s]}) \sim 44.7$, with the protons in particular being closer to equipartition with the magnetic power $\log_{10}(L_{\rm B}\;{\rm[erg/s]}) \sim 45.9$, mainly due to assuming a high $\gamma_{\rm min}$ of the electrons.

The X-ray flux appeared not to change significantly between the two \nustar
observations, and this is supported by a relatively constant (with variability 
less than a factor of 2) hard X-ray flux as seen by Swift BAT data (see, e.g., \cite{Neronov_2015}, 
but also more complete hard X-ray light curve at the NASA's Swift BAT 
site\footnote{\href{https://swift.gsfc.nasa.gov/results/bs157mon/983}{https://swift.gsfc.nasa.gov/results/bs157mon/983}}). We also apply a 
novel method of time-series analysis to the \nustar X-ray data (possible with time-stamps on each X-ray event in both \nustar detectors), 
and detect no short-time variability on hour time scales.  All this implies 
that (1) the X-ray emission is either not co-spatial with, or 
significantly larger in size than the \gm-ray emission region, or 
(2) that the particle acceleration and cooling time scale for the \gm-ray
producing electrons (those radiating above the cooling break) is very rapid, 
while it is significantly longer (than the source crossing time) 
for those responsible for the X-ray flux.  We adopt the latter, simpler 
scenario, implying a magnetic field strength of $B\sim0.04$ G and an electron energy 
distribution extending down to a relatively low Lorentz factor 
$\gamma_{el}$ of a few.  We also argue on the grounds of 
the jet energetics and charge neutrality that the composition 
of the radiating particles is more likely to include a significant fraction 
of positrons, rather than having one proton per electron.


\section{Conclusions}
\label{sec:concl}
In summary, 
we provide a model for the broad-band 
spectral energy distribution of PKS 1830-211 as arising from a combination 
of synchrotron radiation  
(radio-to-optical) and inverse 
Compton scattering 
(X-ray and \gm-ray),
the latter 
by the synchrotron-producing electrons upscattering 
the dusty torus infrared flux. 
We discovered that the X-ray variability 
is of much lower amplitude than 
that of
the large amplitude \gm-ray flares.  

To elucidate lensing delays 
in this source,
we elaborated and applied the well-known 
auto-correlation technique.
This provides 
clear evidence for the presence of a 
non-varying lag at
$t_0=21.1 \pm 0.1\,$d
in the \fermilat 
\gm-ray time series of \pks.

The magnification ratio,
denoted $a$ in Equation (\ref{eq:lc_model})
is less well determined 
and subject to a 
two-fold degeneracy;
however our results are consistent with 
this parameter also being constant 
over the full observation interval.

The methods and results presented 
here have implications in several
contexts.
At other wavelengths 
(for example visible light surveys 
such as the Zwicky Transient Factory,
and future projects 
like the Vera Rubin Observatory)
many resolved lensing  systems 
can be studied.
But in addition, 
monitoring variable 
unresolved images 
will provide many 
gravitational lens
candidates.
Delay detection methods 
like those developed here
could be useful in 
assessing the possibility that 
some of these have arguably composite
light curves and are therefore such candidates.

The unbinned spectral and timing analysis
methods described here can be useful in
extracting scientific information without 
the neglect and degradation of information 
that binning and smoothing entail.
Directly from the 
Fourier transform 
estimated as in 
Section \ref{sec:unbinned_ft} 
and
Appendix \ref{sec:ft_uneven}, 
many other statistical 
quantities can be derived:
power and phase spectra, 
cross-power and cross-phase 
spectra, co-spectra
(limited to data 
from telescopes with 
equivalent dual focal planes,
as with \nustar)
and time resolved versions
of all of these.

We plan to build upon the metric optimization 
scheme (Appendix \ref{sec:MO}).
One can define and test 
an almost infinite variety
of metrics, 
identifying sweet spots for them,
and perhaps 
optimizing combinations of several metrics.
This could lead to methods with
much improved sensitivity to composite 
light curves in the sense of 
combinations of delayed, magnified 
and co-added signals.

Both this and the auto-correlation method can be extended to multiple images which will be explored in future work.


\begin{acknowledgements}
SMW is thankful for her fellowship with the Stiftung der deutschen Wirtschaft (sdw) and acknowledges support through the Baron von Swaine Scholarship by the Universit\"atsbund W\"urzburg. She is grateful for the hospitality and fruitful discussions at Stanford University and SLAC.

\newline SMW, PG, and KM acknowledge funding by the Deutsche Forschungsgemeinschaft (DFG, German Research Foundation), within the research unit 5195 “Relativistic Jets in Active Galaxies” under project number 443220636.

\newline This project was partially supported by the Polish National Science Centre grant 2021/41/B/ST9/04306 as well as NASA GRANT NO.: 80NSSC19K0991.

\newline This research has made use of the NuSTAR Data Analysis Software (NuSTARDAS) jointly developed by the ASI Space Science Data Center (SSDC, Italy) and the California Institute of Technology (Caltech, USA).

\newline The \textit{Fermi} LAT Collaboration acknowledges generous ongoing support from a number of agencies and institutes that have supported both the development and the operation of the LAT as well as scientific data analysis. These include the National Aeronautics and Space Administration and the Department of Energy in the United States, the Commissariat \`a l'Energie Atomique and the Centre National de la Recherche Scientifique / Institut National de Physique Nucl\'eaire et de Physique des Particules in France, the Agenzia Spaziale Italiana and the Istituto Nazionale di Fisica Nucleare in Italy, the Ministry of Education, Culture, Sports, Science and Technology (MEXT), High Energy Accelerator Research Organization (KEK) and Japan Aerospace Exploration Agency (JAXA) in Japan, and the K.~A.~Wallenberg Foundation, the Swedish Research Council and the Swedish National Space Board in Sweden.

Additional support for science analysis during the operations phase is gratefully acknowledged from the Istituto Nazionale di Astrofisica in Italy and the Centre National d'\'Etudes Spatiales in France. This work performed in part under DOE Contract DE-AC02-76SF00515.

\newline This work made use of the following open-source Python packages: 

\href{https://numpy.org}{NumPy} \citep{numpy_Harris_2020},

\href{https://scipy.org}{SciPy} \citep{scipy_2020},

\href{https://matplotlib.org/}{Matplotlib} \citep{matplotlib_Hunter_2007},

\href{https://github.com/felixpatzelt/colorednoise}{colorednoise},

\href{https://pypi.org/project/lightcurves/}{lightcurves} \citep{Wagner_2021},

and \href{http://www.astropy.org}{Astropy}: a community-developed core Python package and an ecosystem of tools and resources for astronomy \citep{astropy_2013, astropy_2018, astropy_2022}.

\end{acknowledgements}

%
%

\bibliographystyle{aa} 
\bibliography{main} 


\begin{appendix}
\label{appendix}

\section{Independent X-ray Spectral Analysis}
\label{sec:xray_spec_independent}
\subsection{\swiftxrt spectrum}
Based on the considered \swiftxrt observations (Tab.\,\ref{tab:swift_obs}), we obtain four spectra that are fitted individually within 0.5 and 10\,keV using an absorbed power-law as described above (\texttt{tbabs$\,\times\,$cflux(powerlaw)}). The results are listed in Tab.\,\ref{tab:swift_spec}. 
The best fit parameters for N$_{\mathrm{H}}$ fixed to $0.188\,\frac{10^{22}}{\text{cm}^{2}}$ are in agreement with results published in the Open Universe for blazars\footnote{Open Universe for blazars, \href{https://sites.google.com/view/ou4blazars}{https://sites.google.com/view/ou4blazars}} \citep{Giommi_2021}, where a similar automated analysis was performed on 82 \swiftxrt observations. However, leaving N$_{\mathrm{H}}$ free to vary yields systematically better fits ($\chi^2_C$/dof closer to one) with larger power-law indices (meaning softer underlying continua). This is expected, as the two fit parameters, the power law index and the fitted absorption, are correlated. The flux levels indicate that the X-ray flux and spectrum of the source of 2016 are very similar to those of 2019. We note that the bandpass of the \swiftxrt instrument is quite limited, and, as we note below, the addition of \nustar data provide a much more complete picture.

\subsection{\nustar spectrum}
In analogy, the spectra of FPMA and FPMB from \nustar are simultaneously fit within the range 3 through 50\,keV with an additional constant factor to account for calibration differences (\texttt{constant$\,\times\,$tbabs$\,\times\,$ cflux(powerlaw)}) for each observation. The normalization of the powerlaw and the constant of FPMA are set to one which results in a constant of 1.04 for FPMB. The fit results are listed in Tab.\,\ref{tab:nustar_spec}.

Both \nustar observations are also listed in the hard X-ray spectroscopic catalogue of blazars \citep[\textit{NuBlazar;}][]{Middei_2022} where similar spectral fits were performed for a large sample of sources. Our results for fixed N$_{\mathrm{H}}$ are in agreement within the error bars with their best-fit values for \pks.
Due to the wider bandwidth of \nustar, it is possible to determine the power-law index more precisely (leading to $\Gamma\approx 1.5$) than with the \swiftxrt data alone. Moreover, the energy range covered by \swiftxrt is more affected by absorption. This shows that particularly for such sources, \nustar -- with the bandpass extending to higher energies, where the observed spectrum is much less affected by photoelectric absorption -- is crucial for providing data to adequately constrain spectral properties. Combination of the two instruments yields (quasi-)simultaneous data which allows us to constrain the absorption as well as the power-law index more rigorously.

\begin{table}[]
    \centering
    \caption{\swiftxrt spectral analysis best fit parameters for \texttt{tbabs$\,\times\,$cflux(powerlaw)} with the power-law index $\Gamma$ and the Galactic H$_\text{I}$ column density N$_{\mathrm{H}}$ free to vary (top rows) and fixed to the Galactic value (bottom rows) as well as the estimated flux $F_{0.5-10}$ within 0.5 and 10\,keV in $10^{-12}$ erg/(cm$^2$s) for the 2016 and 2019 data (see left column).}
    \resizebox{\linewidth}{!}{
    \begin{tabular}{c|c|c|c|c|c}
        yy & ID & N$_{\mathrm{H}}$ $\left[\frac{10^{22}}{\text{cm}^{2}}\right]$ & $\Gamma$ & $\chi^2_C$/dof & $F_{0.5-10}$ $\left[10^{-12} \frac{erg}{cm^2s}\right]$ \\ 
        \hline
        \hline
        16 & 02 & $0.67 \pm 0.17$ & $1.13 \pm 0.19 $ & 0.98 & $ 19.9 \pm 1.4 $ \\ \hline 
        16 & 16 & $0.85 \pm 0.13$ & $1.20 \pm 0.12 $ & 0.92 & $ 20.47 \pm 0.93 $ \\ \hline 
        19 & 36 & $0.88 \pm 0.19$ & $1.42 \pm 0.19 $ & 0.75 & $ 21.4 \pm 1.6 $ \\ \hline 
        19 & 37 & $0.94 \pm 0.23$ & $1.23 \pm 0.20 $ & 1.11 & $ 21.3 \pm 1.6 $ \\ \hline 
        \hline
        16 & 02 & $0.188$ & $0.65 \pm 0.11 $ & 1.08 & $ 19.9 \pm 1.7 $ \\ \hline 
        16 & 16 & $0.188$ & $0.558 \pm 0.068 $ & 1.17 & $ 20.3 \pm 1.1 $ \\ \hline 
        19 & 36 & $0.188$ & $0.73 \pm 0.10 $ & 0.99 & $ 20.3 \pm 1.6 $ \\ \hline 
        19 & 37 & $0.188$ & $0.56 \pm 0.11 $ & 1.31 & $ 20.8 \pm 1.8 $ \\  
    \end{tabular}
    }
    \label{tab:swift_spec}
\end{table}

\begin{table}[]
    \centering
    \caption{\nustar spectral analysis best fit parameters for \texttt{constant$\,\times\,$tbabs$\,\times\,$ cflux(powerlaw)} with the power-law index $\Gamma$ and the Galactic H$_\text{I}$ column density $n_H$ free to vary (top rows) and fixed to the Galactic value (bottom rows) as well as the estimated flux $F_{3-50}$ within 3 and 50\,keV in $10^{-12}$ erg/(cm$^2$s) for the 2016 and 2019 data (see left column).}
    \resizebox{\linewidth}{!}{
    \begin{tabular}{c|c|c|c|c|c}
        yy & ID & N$_{\mathrm{H}}$ $\left[\frac{10^{22}}{\text{cm}^{2}}\right]$ & $\Gamma$ & $\chi^2_C$/dof & $F_{3-50}$ $\left[10^{-12} \frac{erg}{cm^2s}\right]$ \\
        \hline
        \hline
        16 & 6016 & $1.14 \pm 0.53$ & $1.477 \pm 0.025 $ & 0.93 & $ 49.55 \pm 0.79 $ \\ \hline 
        19 & 8046 & $0.53 \pm 0.39$ & $1.450 \pm 0.019 $ & 0.92 & $ 46.71 \pm 0.56 $ \\ \hline 
        \hline
        16 & 6016 & $0.188$ & $1.441 \pm 0.015 $ & 0.94 & $ 49.61 \pm 0.80 $ \\ \hline 
        19 & 8046 & $0.188$ & $1.437 \pm 0.011 $ & 0.92 & $ 46.74 \pm 0.56 $ \\ 
    \end{tabular}
    }
    \label{tab:nustar_spec}
\end{table}

\section{Unbinned photon analysis}

\subsection{The Frequency Response Function}
\label{sec:ft_uneven}
Of considerable interest is the frequency response function, or frequency-domain point spread function (PSF), which is unavoidably convolved with the true power spectrum. It is important because it thus determines the ultimate spectral resolution of the estimated power spectrum. Its importance, even for ordinary spectral analysis of evenly sampled data, is often overlooked -- for example in its role in 
various forms of \textit{spectral leakage}. The PSF describes how power leaks away from the true frequency into \textit{sidelobes} of order $1/T$ c/sec in width. Its shape depends on the sampling of the data in time. Hence a possible concern is that the irregular nature of the sampling in this unbinned algorithm may yield peculiar sidelobes. 

\begin{figure}[h]
\begin{center}
      \includegraphics[width=0.8\linewidth]{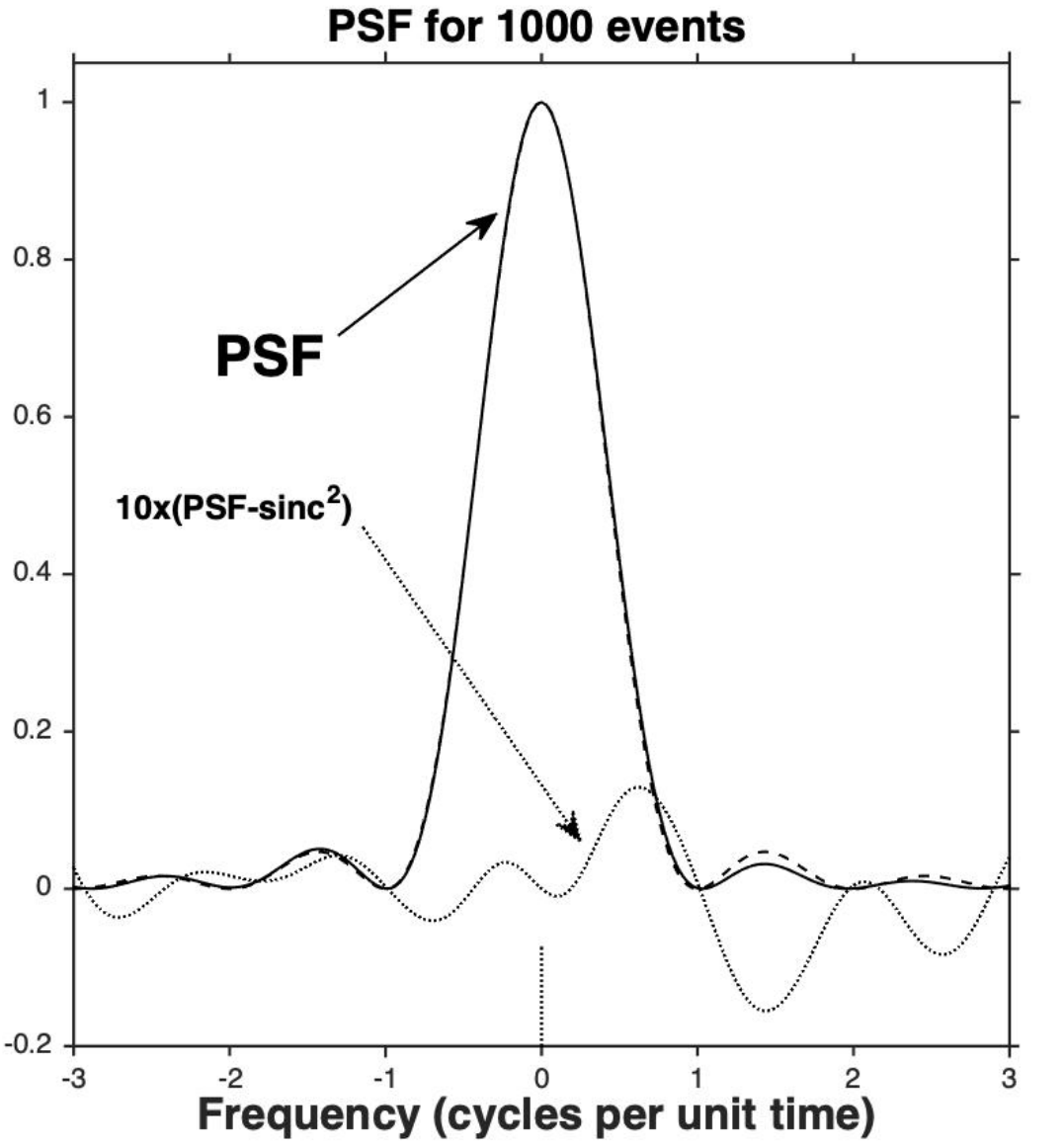}
    \caption{Spectral response function (solid line) as described in the text, for 1000 simulated unbinned event times. For more than this number the difference from the sinc$^2$ profile rapidly becomes insignificant; for much less the sidelobes become quite different, as expected. The  data interval $T$ is taken to be $1$, so the fundamental frequency is also unity. The fiducial frequency $f_0$ in Eq. (\ref{sin_prob}) was taken as 8,000 times the fundamental. The dashed line, largely overlaid by the solid curve, is 
    the ideal "sinc-squared" sidelobe function $sin^{2}( f ) / f^{2}$. The dotted line is 10 times the difference between these curves.
    }
    \label{fig:psf_1}  
\end{center}    
\end{figure}

Accordingly, in Fig.\,\ref{fig:psf_1} we address this specific question: 
How does this response function compare to the ideal response function for evenly spaced data on the same interval? 
As we now demonstrate, the answer in brief is: Very accurately they are the same; the PSF sidelobe has no significant peculiarities.

To compute the spectral PSF we can't simply replace the sampled values $x_{n}$ with unity, which is the usual way to compute the \textit{window function} for arbitrarily sampled data; with the current algorithm that just reproduces the data's power spectrum. So we fall back on the definition and compute the power spectrum of a purely monochromatic signal -- a sinusoid. We need a synthetic signal corresponding to event times randomly drawn from a probability distribution that has a sinusoidal time dependence. To do this we employed the standard trick of uniform random draws in the cumulative distribution function derived from a differential probability distribution of the form 

\begin{linenomath*}
\begin{equation}
    p(t) = 1 + \mbox{sin} ( f_{0} t ) \ \ ,
    \label{sin_prob}
\end{equation}
\end{linenomath*}
\noindent
evaluated at an ensemble of random times. 
$f_0$ is a fiducial frequency, the value of which does not change the results as long as it is much larger than the fundamental. The offset of $1$ is to avoid negative probabilities, and is irrelevant to the power spectrum at frequencies other than zero. The main dependence is on the number of simulated events. The mean departure from the ``sinc-squared'' function is negligible for a typical GTI of the \nustar data analyzed here, as suggested by the comparison shown in Fig.\,\ref{fig:psf_1} for the comparatively small sample of 1000 events. The small asymmetry seen in the PSF, relative to the exactly symmetric sinc$^2$ function, is due to the luck of the draw in the data points in the particular realization depicted in the figure. The shape of the PSF depends on the point distribution on the scale of the full observation interval, not on the fine scales associated with mean inter-point intervals.

\noindent

\subsection{Application to \fermilat data of \pks}
\begin{figure}[h]
    \includegraphics[width=\linewidth]{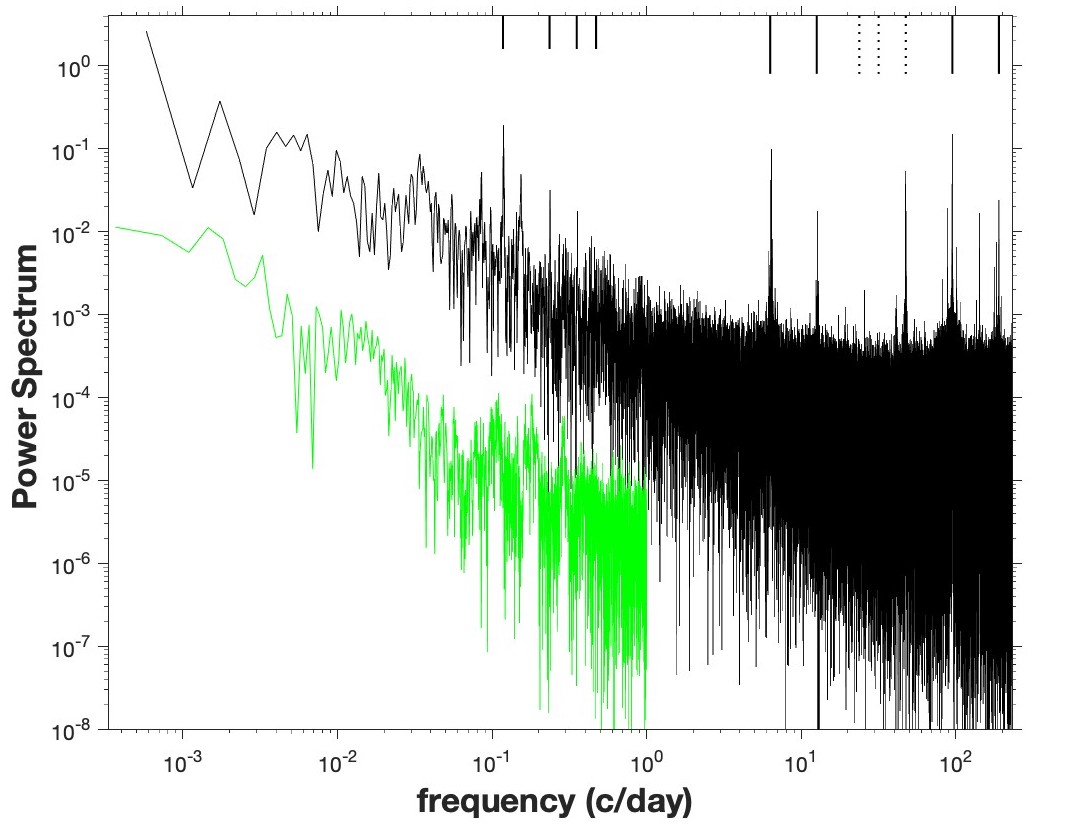}
    \caption{Power spectra for \pks.
    Black curve:  unbinned power 
    spectrum, 
    based on 209,206
    photons,
    weighted 
    according to 
    their distance 
    from the source center
    by a Gaussian
    with a 1 degree
    standard deviation.
    The detected instrumental 
    modulation frequencies are 
    indicated  by vertical 
    lines at the top of the
    panel; dotted lines are
    subharmonics.
    Green curve: FFT power
    spectrum of the daily 
    samples discussed above,
    offset vertically
    by a factor of
    100 for clarity.
    }
    \label{fig:photon_sp}
\end{figure}

Figure \ref{fig:photon_sp}
demonstrates 
the effectiveness of this
unbinned 
power spectrum estimator.
Individual photon
weights were 
assigned using the 
focal plane 
distance $d_n$
of the photon 
from the source
position:
\begin{linenomath*}
\begin{equation}
    w_{n} \sim  
    e^{- r_{n} ^2 /
    \sigma ^2 } \ , 
\end{equation}
\end{linenomath*}
with 
the choice $\sigma = 1$ 
giving good 
background discrimination.

Because it derives 
information from photons
arbitrarily close in time,
it provides 
an estimate of high frequency
power that is simply 
not accessible to the
Fourier spectrum of 
the daily samples.
In addition to detecting
the Fermi spacecraft precession
frequency 
(a period of 53.4 days)
and its first three harmonics,
it detects modulations 
at both 1 sidereal day 
(86164.0905 seconds)
due to the effect of the
South Atlantic Anomaly 
on data acquisition, 
and the orbital period
of 95.2 minutes.
Not only are harmonics
of the orbit period 
clearly visible,
but they are straddled 
by a series of beat
frequencies with the 
sidereal day modulation.
In turn, the
sidereal day peak 
exhibits beat frequencies
with the precession modulation.

Of course these 
periodic signals 
have nothing to do 
with the variation of
\pks, but can be thought 
of as contamination 
that could be removed 
by filtering, if desired.
While there is no
peak 
in Figure \ref{fig:envelope}
at the precession 
period, 
the one at about 
35 days 
is close to 
a potential 
beating of 
precession 
against the 22 day lag.

\noindent

\section{Modeling Lensed Light Curves}
\label{appendix:lensing_model}

Using the fact that the Fourier transform of a signal $x(t - t_{0})$, i.e. a copy of a signal $x(t)$  delayed by $t_{0}$, is $e^{-i\omega t_{0} }$ times the transform of the undelayed signal, the Fourier transform of the observed signal in Eq. (\ref{eq:lc_model}) can be expressed as
\begin{linenomath*}
\begin{equation}
        Y(\omega) = X(\omega)\, (1 + a \, e^{-i \omega t_0 \ } )
\label{eq:lens_trans}
\end{equation}
\end{linenomath*}
Solving 
for $X( \omega )$ 
as a function of $Y( \omega )$,
and then inverse transforming yields
\begin{linenomath*}
\begin{equation}
\boxed{
 x( t )  = IFT \left[  
    { Y(\omega) \over 
    1 + a e^{-i\omega t_{0} } } \right] .
    }
    \label{eq:mo_x}
\end{equation}
\end{linenomath*}
This is an exact solution 
for the intrinsic light curve, 
given the observed light curve and
values of 
$t_{0}$ and $a$.
This is the key result 
in estimating these parameters 
in the Metric Optimization approach.\\

This can be used to derive corresponding
spectrum 
relation, and to show that 
the power spectrum $P_{Y}(\omega) = |Y(\omega)|^{2}$ is 
modified from that for the intrinsic light curve 
by a relatively simple 
frequency-dependent factor, viz. 
\begin{linenomath*}
\begin{equation}
        P_{Y}(\omega) = P_{X}(\omega) \ (1 + 2 a \,\text{cos}(2 \pi  \omega t_0 ) + a^2 ) .
\label{eq:power_spectrum_solution}
\end{equation}
\end{linenomath*}
This relation exhibits an interesting property: \textit{the power spectrum of the observed light curve is a sinusoidally modified version 
of the power spectrum of the intrinsic light curve.} Thus, the time delay can be approached by 
searching for a peak in the 
power spectrum of the power spectrum
or in the cepstrum\footnote{This 
modification of the power spectrum
was developed to study 
``echoes'' exactly as in  
equation (\ref{eq:lc_model} \citep{Bogert_1963})}.
We do not pursue this concept here, 
but the ``double power spectrum'' method 
developed and applied by \cite{Barnacka_2011,Barnacka_2015}
is explicitly based on this result.

A related approach to the delay problem 
was developed several decades ago, principally by
John Tukey 
\citep{Bogert_1963}.
Exactly equation (\ref{eq:lc_model}) 
and our solution 
in equation (\ref{eq:power_spectrum_solution})
are studied in this reference,
in the context of detecting 
``echoes'' e.g. in radar research.
In particular, 
these authors defined the 
\textit{cepstrum}\footnote{
Their terminology for the 
variables parallel to 
frequency (quefrency).
phase (saphe),
analysis (alanysis), 
etc.
seem quaint today.
Other than the term \textit{cepstrum} for 
the variant of the ordinary spectrum,
these 
terms have largely disappeared.}
as the (Fourier) power spectrum 
of the logarithm of the (Fourier)
power spectrum of the original 
process.\\

The auto-correlation function (ACF) for evenly sampled data is defined as 
\begin{linenomath*}
\begin{equation}
R_{Y}( \tau ) 
= {1 \over N } \  \sum\limits^N_{n=0} \ y( t_{n} ) \  y( t_{n} - \tau ) \ .
\label{acf_def}
\end{equation}
\end{linenomath*}
This expression does not include
three commonly imposed procedures:
normalization by the variance,
adjustment for bias, 
or adjusting $y$ by subtracting off its mean value.

Using Equation (\ref{eq:lc_model}), 
$R(\tau)$ is represented
as the average over $t$ of
\begin{linenomath*}
\begin{equation}
y(t) y( t - \tau) = 
   [ x(t) + a x( t - t_{0} ) ]
   [ x(t - \tau) + a x( t - \tau- t_{0} ) ] \ .
\label{eq:acf_1}
\end{equation}
\end{linenomath*}
Expanding this expression, and
noting that a shift in $t$ 
yields an offset in the argument $\tau$ of the ACF, we have
\begin{equation}
R_{Y}(\tau) = 
( 1 + a^{2} )  R_{X}(\tau )
+
a [ R_{X}(\tau + t_{0} ) +  R_{X}(\tau - t_{0} ) ] 
\label{eq:acf_model_append}
\end{equation}
\noindent
Since these three terms 
overlap each other,
care must be taken in 
interpreting the last two 
as ``bumps'' superimposed 
on the first.\\

The generalization of Eq.~(\ref{eq:lc_model})
to an arbitrary number $N_{paths}$ 
of lensing components 
can be computed analogously \citep{Wagner_phd} but are omitted here due to the page limit.

\section{Autocorrelation Functions of Data with Gaps}
\label{sec:gaps}

Regarding real world measurements,
the estimation of the Auto-correlation function is slightly
complicated by the presence of gaps 
in the otherwise evenly sampled light curves.
Flux estimates are missing whenever the
source flux and/or the instrumental sensitivity 
are too low to yield more than an upper limit.

For the autocorrelation function, for example,
we considered four different ways of 
dealing with this issue, the first two 
explicitly insert values at the 
times of the missing data,
the second two avoiding these unsampled times:
\begin{enumerate}
\item Linearly interpolate the good values into the gaps
\item Replace the missing points with a quiescent flux value (e.g. the median flux)
\item Use the DCF algorithm of \cite{EdelsonKrolik_1988}
\item Use the inverse Fourier transform of the Lomb-Scargle periodogram (LSP)
\end{enumerate}
The first three implement the concept embodied in Eq. (\ref{acf_def}),
namely the inner product of the time series
with a shifted version of itself.
Number 4 makes use of 
a well-known theorem \citep{Bracewell_FT} 
expressing the autocorrelation function 
as the inverse Fourier transform  of the power spectrum:
\begin{linenomath*}
    \begin{equation}
    \label{eq:autocorr-theorem}
        R_y(\tau) \equiv \int y(t)\, y(t+\tau) \, dt  =  \int |Y_{LS}(\omega)|^2 e^{\ i 2\pi \omega \tau} \, d\omega
    \end{equation}
\end{linenomath*}
and estimating the LS power spectrum
with Eq. (\ref{lomb_scargle}).

\begin{figure}
    \centering
    \includegraphics[width=\linewidth]{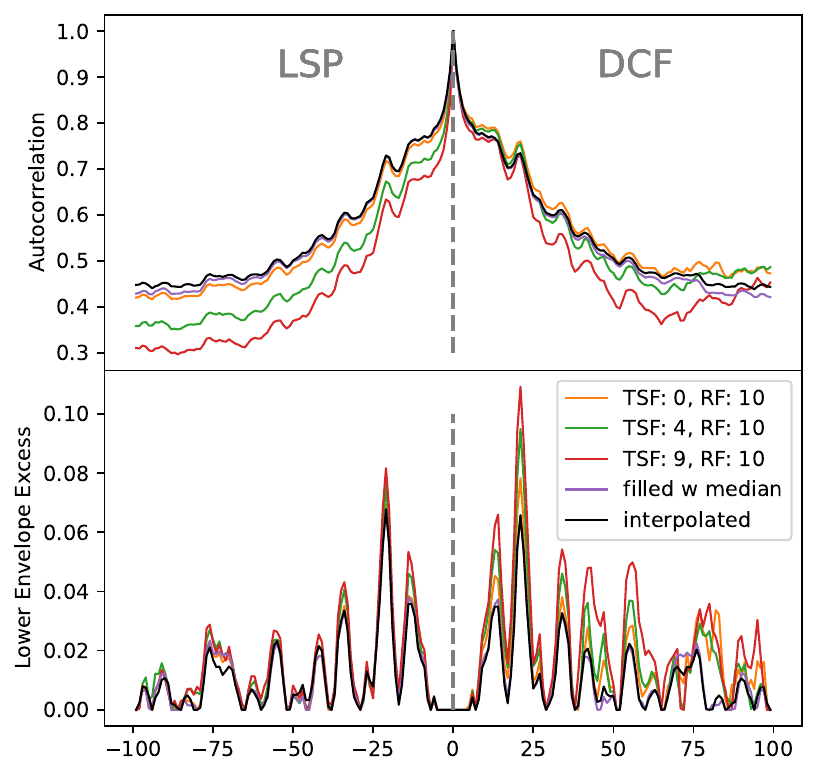}
    \caption{\textit{Top}: Autocorrelations obtained with the Fourier Transform of the LSP (left) and the DCF (right) normalized to 1 at zero lag. \textit{Bottom}: Excess of the lower envelope (see Sect.\,\ref{sec:lag}). The colors correspond to different choices of data quality/dealing with gaps (see legend). For this work, we used interpolation for gaps in the light curve yielding the black solid lines.}
    \label{fig:acf_methods}
\end{figure}

We illustrate the ACFs obtained with these methods in the top panel of  Fig.\,\ref{fig:acf_methods} along with the corresponding lower envelope excess (see Sec.\,\ref{sec:lag}) in the lower panel.
The left panel of the figure shows computation with the inverse Fourier transform of the LSP and the right panel depicts the DCF.
Different lines indicate different handling of the light curve data.
The orange, green, and red lines correspond to threshold values of 0, 4, and 9 for the test statistic filter (TSF, see Sec.\,\ref{sec:data_fermi}).
The black line was obtained through interpolation of gaps, while the purple line results from replacing gaps with the median flux, corresponding to method 1 and 2.

LSP and DCF are attractive, commonly used methods 
for dealing with missing data.
They implicitly 
make the assumption that the missing points
occur randomly 
and independently 
in time.  
For \fermilat the absence of data is not completely
random but rather largely systematic,
occurring 
when either the true flux or 
the instrumental live-time or exposure
are so low that the source
does not rise above the 
detection threshold.
Accordingly these 
methods can introduce a statistical bias.

Method 2 (replacing gaps with the median flux) may be reasonable 
during quiescent times, 
but will distort the light curve 
when the dropout is due to low 
instrumental live time rather
than low source flux.
The remaining method, interpolation,
does introduce unobserved data in place of the
missing points, but is clearly better than
method 2 in this regard.

The overall shape of all ACFs (as in the sizes and locations of the various bumps) and particularly the lower envelope excess are closely the same in all cases.
\cite{Barnacka_2015} 
similarly found that two other treatments 
of upper limits (using zero or the
upper limit itself as the flux value) 
yielded negligible differences 
in estimated time delays and their significances.
Those authors also used interpolation
for points of low test statistic, and 
illustrated in some detail the dependence of the significance of peaks in the ACF on the nature of source variability.

Taking all of these factors into account,
we have adopted the LSP method for the atuo-correlation approach and linear interpolation for the Metric Optimization approach reported in the main body.

\section{Bootstrap Error Estimates}
\label{sec:bootstrap}

To assess the statistical uncertainties 
of these parameter estimates,
we used a bootstrap randomization procedure.
Time series data cannot 
be directly bootstrapped 
using the standard procedure
\citep{Efron_1994}.
This is because 
flux samples 
lack the statistical independence
crucial to the bootstrap's
\textit{random drawing with replacement}
procedure.
On the other hand, 
the \gm-ray photons themselves 
are independent: arrival of a 
photon at one time does not influence the 
probability of arrival of photons at
any other time.\footnote{Astronomical
flux variation is  typically what is 
known in statistics as a Cox, or 
\textit{doubly stochastic process}, 
so-named because it 
consists of two separate random processes:
photon detection and source variability \citep{Snyder_2012}.
Absent correlated observational noise, 
the former is an independently distributed,
and therefore white, process.
The latter is typically dependently distributed,
with correlations and memory on various scales.
}
Re-sampling photon times is a valid bootstrap procedure, as long as 
duplicate time-tags are accounted for properly. 

Each flux value 
(photons per cm$^2$ per sec) 
corresponds to an expected number of photons.
We used a rough estimate of this 
conversion factor -- 
Flux $= C N_{photons}$  -- 
by 
assuming a Poisson photon distribution
(not strictly true, but
approximate)
for which 
\begin{linenomath*}
\begin{equation}
\sigma( F ) = C \sigma(N_{photons}) = C \sqrt N_{photons}  
\label{eq:poisson}
\end{equation}
\end{linenomath*}
and simple algebra gives 
\begin{linenomath*}
\begin{equation}
{N_{photons} \over F} = {F \over \sigma^{2}(F)}
\label{eq:poisson_1}
\end{equation}
\end{linenomath*}
thus allowing an estimate of $N_{photons}$
by multiplying each flux sample by
an average value of the ratio on 
the right hand side of this equation.
This method neglects 
the added uncertainty due to background,
which has already been 
accounted for in determining 
these fluxes.
Thus we construct a random sequence 
of simulated photon times 
for each flux sample;
these times can then be 
bootstrapped directly.
The resulting uncertainties are 
reported in the tables and 
plots of this paper.

\section{Extraction of Magnification Ratio from ACF}
\label{sec:extract_a}
Once the value of the lag $t_{0}$ has been 
determined, the value of $a$ 
can be found  from 
Eq. (\ref{eq:acf_model}) evaluated 
at lag zero 
(making use of the  symmetry of $R_{X}$)
\begin{linenomath*}
\begin{equation}
R_{Y}(0) = 
( 1 + a^{2} ) R_{X}(0)
+ 2 a R_{X}(t_{0}) 
\label{acf_model_zero}
\end{equation}
\end{linenomath*}
\noindent 
and at lag $t_{0}$:
\begin{linenomath*}
\begin{equation}
R_{Y}(t_{0}) = 
( 1 + a^{2} ) R_{X}(t_{0})
+ a [ R_{X}(2t_{0}) + R_{X}(0) ]
\label{acf_model_t0}
\end{equation}
\end{linenomath*}
\noindent 

The goal now is to obtain 
an equation for $a$
in terms of known
quantities  -- 
such as 
the left-hand sides 
of the above two equations,
obtained directly from 
the ACF of the data.
To accomplish this we 
estimate 
the unknown coefficients 
involving $R_{X}$
on the right-hand sides.
It turns out that 
only the
relative values 
of $R_{X}$ at the three 
lags $0, t_{0},$
and
$2 t_{0}$
are needed.
These
dimensionless ratios 
are determined by the
shape of $R_{X}(\tau)$,
independent of its 
normalization.
Now note that
Eq. (\ref{eq:acf_model})
implies that 
the lower envelope 
of $R_{Y}(\tau)$ 
is a good estimate of 
its first term -- i.e.
the shape of 
$R_{Y}(\tau)$ with 
the bumps corresponding
to the last two terms
removed.
Accordingly,
the ratios mentioned above
can be derived from the 
shape of the lower envelope
of
$R_{Y}(\tau)$\,
using the fact 
that the shape of
$R_{X}(\tau)$ 
and that of the 
first term in 
Eq. (\ref{eq:acf_model})
are the same.

Specifically,
denoting the lower envelope
of $R_{Y}(\tau)$ by $L(\tau)$,
define two ratios
\begin{equation}
    F_{1} = L(t_{0} ) / L( 0 )
\end{equation}
and
\begin{equation}
    F_{2} = L(2 t_{0} ) / L( 0 ) \ .
\end{equation}
These constants can be determined directly from the
lower envelope 
as plotted in 
Figure \ref{fig:envelope}.
Furthermore the 
assumption that
the shapes of 
$L(\tau)$ and $R_{X}(\tau)$
are the same means 
that 
\begin{equation}
R_{X}(t_{0}) = F_{1} R_{X}( 0 )
\end{equation}
and
\begin{equation}
R_{X}(2 t_{0}) = F_{2} R_{X}( 0 ) \ .
\end{equation}
\noindent 
Now we can write equations (\ref{acf_model_zero}) 
and
(\ref{acf_model_t0}) as
\begin{equation}
R_{Y}(0) = 
( 1 + a^{2} + 2 a F_{1} )\, R_{X}(0)
\label{acf_model_zero_f}
\end{equation}
\noindent 
and
\begin{equation}
R_{Y}(t_{0})= 
[( 1 + a^{2} ) F_{1} 
+ a ( F_{2} + 1 ) ] \, R_{X}(0) 
\label{acf_model_t0_f}
\end{equation}
\noindent 
so that we have the ratio 
\begin{equation}
{R_{Y}(t_{0}) \over R_{Y}(0)} = 
{
( 1 + a^{2} ) F_{1} 
+ a ( F_{2} + 1 )  
\over
( 1 + a^{2} + 2 a F_{1} ) }
\label{acf_model_t0_f}
\end{equation}
yielding a linear equation 
in 
$    A = a / ( 1 + a^{2} ) \ ,$
namely  
\begin{equation}
( 1 + 2 A F_{1} )  
{R_{Y}(t_{0}) \over R_{Y}(0)} = 
F_{1} + A ( F_{2} + 1 ) 
\label{acf_model_t0_f}
\end{equation}
so that
with $R= { R_{Y}(t_{0}) 
  \over 
  R_{Y}(0)}$
we have
\begin{equation}
A = 
{  
   R - F_{1}
\over
  F_{2} + 1 
  - 2 F_{1} R
}
\label{acf_model_t0_f}
\end{equation}
\noindent
which from the relation above 
can be plugged into 
\begin{equation}
    a = {1 \over 2A}  \pm 
    \sqrt{ ( {1 \over 2A})^{2} - 1 } \ . 
    \label{eq:aa_acf_solution}
\end{equation}
\noindent
It is easy to verify that 
these two solutions 
are reciprocals 
of each other,
as is necessary 
because of the 
time-reversal
invariance of the 
autocorrelation function.
The model 
in Equation (\ref{eq:lc_model}) 
implies that reversing
time effectively
interchanges the 
identities of the 
two lensed paths,
corresponding to 
the relation
$t \rightarrow -t
\Longleftrightarrow
 a \rightarrow 1/a $.

The last two equations 
give our solution 
for the magnification
factor.
But there is one issue to be
addressed before values 
for the needed quantities
$R$, $F_1$ and $F_2$
are obtained 
from the empirical 
autocorrelation function 
$R_{Y}(\tau)$:
it is helpful to reduce or 
remove the effects of 
the errors of observation 
on the ACF, 
ideally leaving an unbiased 
estimate of the autocorrelation 
of the true source variability.

Toward this goal we
consider separately 
errors of observation 
(here "noise")
and intrinsic source 
flux (here "signal"),
and assume 
(1) the noise is 
uncorrelated;
(2) the signal is 
smooth at short time scales;
and
(3) the ACF is the sum of 
two corresponding parts:
\begin{linenomath*}
\begin{equation}
R_{Y}(\tau) =
R_{Y}^{source}(\tau) 
+
R_{Y}^{noise}(\tau)
\ .
    \label{eq:source_noise}
\end{equation}
\end{linenomath*}
\noindent
If the errors of observation 
at different times are uncorrelated
with each other, 
as is characteristic 
of photon count data 
(absent significant 
dead time or pileup effects,
as we assume here)
the noise term 
contributes at $\tau = 0$ 
only -- i.e. it 
is proportional to 
a delta function there.
To achieve a rigorous 
separation
it is necessary to 
make assumptions about 
the source variability.
We propose this estimate
for the source term 
at $\tau = 0$:
\begin{linenomath*}
\begin{equation}
R_{Y}(0) = 
\lim_{\tau \to 0}
R_{Y}^{source}(\tau)
+
\sigma^{2} \ .
    \label{eq:zero_limit}
\end{equation}
\end{linenomath*}
\noindent
which relies on two
assumptions: the 
true autocorrelation 
is well enough 
behaved for 
lags of a few integer
multiples of the observing cadence (1 day)
that a parametric 
extrapolation down to 
zero lag is reasonably
accurate, and that 
the source variability 
does not have a 
significant unresolved
component 
(i.e. at time scales
$\le$ one day)
which would produce 
its own delta function
spike at $\tau = 0$.
The separation condition 
of Equation (\ref{eq:source_noise})
is justified in a wide variety
of noise distributions.
For Gaussian white noise
Equation (\ref{eq:zero_limit})
holds with $\sigma$ the 
dispersion constant 
in the normal distribution.
For pure Poisson noise,
which is not even additive,
standard textbook results give 
\begin{linenomath*}
\begin{equation}
R_{Poisson}(0) = 
\lambda ^{2} 
+
\lambda \ ,
    \label{eq:zero_limit_poisson}
\end{equation}
\end{linenomath*}
\noindent
where $\lambda$ 
is the mean
photon event rate
(cf. Appendix\,\ref{sec:bootstrap}).

\begin{figure}[t]
\centering
\includegraphics[width=\linewidth]
{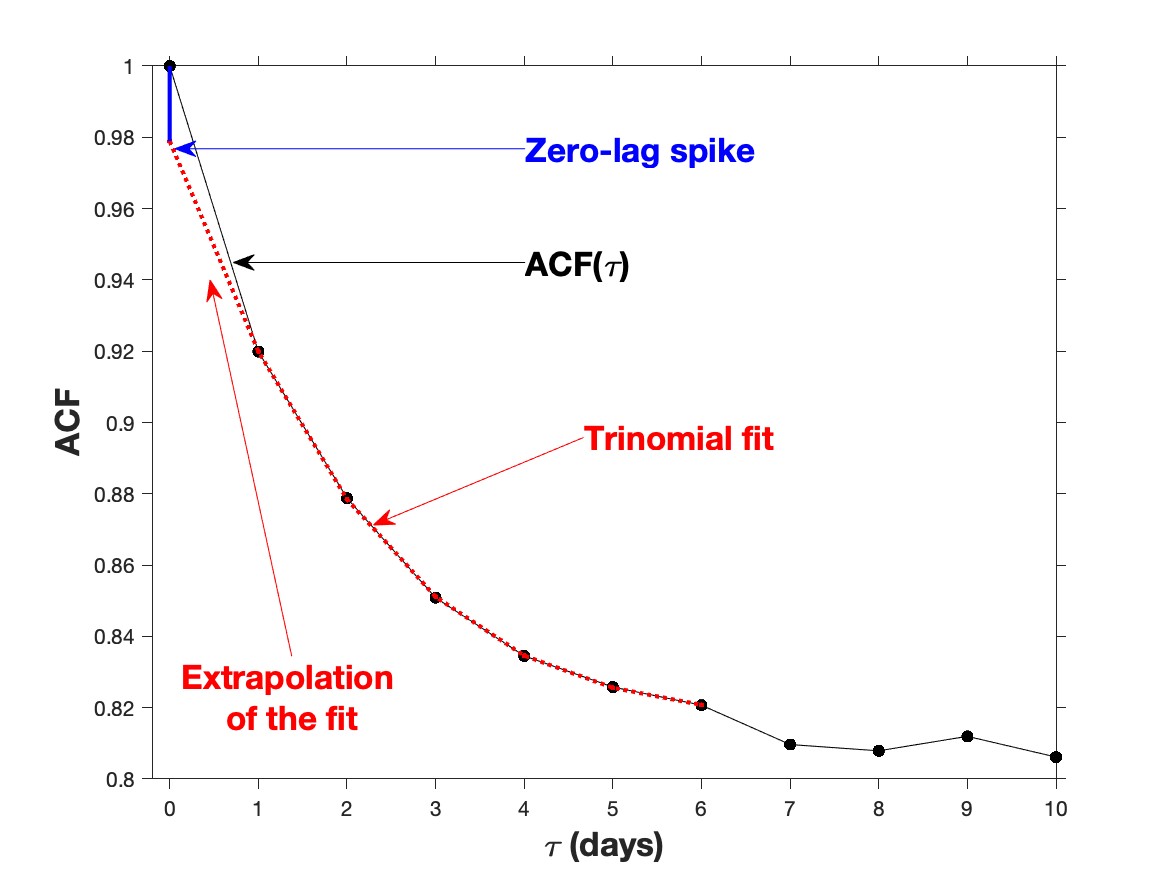}
\caption{Zooming in on the
ACF for \pks 
as shown in Figure~\ref{fig:envelope}.
A polynomial of 3rd degree 
is fit to the ACF at lags
1 to 6, and extrapolated
to lag zero:
the excess there is 
ascribed to the observational
errors and subtracted off 
to form an estimate of the
true ACF at zero lag.
} 
\label{fig:acf_spike}
\end{figure}

In summary:
the ACF possesses 
a delta-function spike 
at zero lag;
using the smoothness 
assumption (2) above, 
it can be removed 
with the simple
extrapolation 
depicted in 
Figure \ref{fig:acf_spike}. 
This procedure yields 
an estimate of the 
correction due to the spike 
that is 2.1\% of the
adopted value of
the true zero-lag ACF.
(The presence of 
an unresolved 
source variation 
would yield an additional 
contribution to $R_{Y}(0)$, 
making 
the already small correction 
even smaller;
we ignore this possibility.)

\section{The Metric Optimization (MO) Procedure}
\label{sec:MO}

A number of approaches to detecting
and measuring time delays and magnification 
ratios in gravitationally lensed AGN have been developed.
\cite{Geiger_Schneider_1996} used an approach similar to the one 
proposed here, 
but differing in two major ways.
Their reconstructed intrinsic light curve $x(t)$ is obtained with an approximate iterative procedure (in comparison to our exact solution, c.f. Equation (\ref{eq:mo_x})). The concomitant issues of convergence, reference to unobserved data past the ends of the observation interval, and accuracy are nicely discussed in their paper. Secondly, they invoke additional (interferometric) data to break the degeneracy in any direct light curve reconstruction (in comparison to our use of physically motivated constraints on  $x(t)$).
A method closer to ours, 
differing mainly in the method of reconstructing $x(t)$, 
was developed  by \cite{Bag_2022B}.
These authors used 
an iterative approach similar to that of \citep{Geiger_Schneider_1996}, 
but resolved the parameter degeneracy by optimizing a metric 
much as we suggest here.

The main difference from 
these previous
approaches 
is use of the solution of the
basic Equation (\ref{eq:lc_model})
for the intrinsic light curve $x(t)$,
derived in Appendix \ref{appendix:lensing_model} as Equation (\ref{eq:mo_x}), and restated here:
\begin{linenomath*}
\begin{equation}
x( t )  =  
IFT \left\{ 
{ FT[ y(t) ] \over 
1 + a \ e^{- i \omega t_{0}} } \right\} \  .
\label{eq:lens_solution}
\end{equation}
\end{linenomath*}
(as well as consideration of a variety of 
quality \textit{metrics}).

This solution for the intrinsic source light curve $x(t)$ in terms of the observed one $y(t)$ and the lens observables ($t_0$ and $a$) 
is exact,
therefore the 
usual optimization of residuals 
from the data cannot be used.
Instead, we propose 
optimizing a quantity 
characterizing some 
quality of 
the solution $x(t)$ -- 
one that 
quantifies 
a property which the composite 
light curve 
$\sum _{n=1}^{N_{paths}}
    a_{n} x( t - t_{n} )$
    has, but the decomposed
    function $x(t)$ does not.
This prescription, 
for what measures a "good" decomposition
of the light curve into 
a linear combination of 
two (or more) components that 
are mutually shifted versions
of each other,
is quite vague.
It is clear that there is no unique best metric for this problem.

We considered a 
large number of 
metrics of $x(t)$.
Many are based on 
its variance, 
or that of it absolute value,
or various differences.
Others measured the
number of peaks (local maxima
in $x(t)$ or its Bayesian 
Block representation.
Some leveraged total variation
in various forms.
The \textit{cepstrum} is a 
modification of the power spectrum
developed to study 
``echoes'' exactly as in  
equation (\ref{eq:lc_model}) 
by \cite{Bogert_1963},
and should be considered 
in this context.
The amplitude of the 
autocorrelation
function at the putative lag,
various information measures, 
etc. 
are also potential metrics.

We emphasize that 
this method is still 
in development,
and further study 
can identify more
efficient metrics,
and measure their 
performance with simulations.

It is straightforward to 
apply 
the same procedure 
to multiple-path lenses,
but in practice the 
optimization is 
of course more difficult.

\end{appendix}


\end{document}